\documentclass[%
 aip,
 amsmath,amssymb,
 reprint,%
]{revtex4-1}

\usepackage{graphicx}
\usepackage{dcolumn}
\usepackage{bm}

\usepackage[utf8]{inputenc}
\usepackage[T1]{fontenc}
\usepackage{mathptmx}
\usepackage{etoolbox}
\usepackage{pstricks}

\graphicspath{{figures/}}

\makeatletter
\def\@email#1#2{%
 \endgroup
 \patchcmd{\titleblock@produce}
  {\frontmatter@RRAPformat}
  {\frontmatter@RRAPformat{\produce@RRAP{*#1\href{mailto:#2}{#2}}}\frontmatter@RRAPformat}
  {}{}
}%
\makeatother

\newrgbcolor{dred}{.67 0 0}

\begin{document}

\preprint{AIP/123-QED}

\title[]{Poly-liquid behaviors of self-associating fluids and mesoscopic aggregation in liquid solutions}
\author{Ilya Davydov}
\affiliation{Department of Chemistry, University of
  Houston, Houston, TX 77204-5003}%

\author{Vassiliy Lubchenko*}%
 \email{vas@uh.edu}
\affiliation{Department of Chemistry, University of
  Houston, Houston, TX 77204-5003}%
\affiliation{Department of Physics, University of Houston, Houston, TX
  77204-5005} \affiliation{Texas Center for Superconductivity,
  University of Houston, Houston, TX 77204-5002}
  
\date{\today}

\begin{abstract}

In conflict with standard notions of thermodynamics, mesoscopically-sized inclusions (``clusters'') of a solute-rich liquid have been observed in equilibrated solutions of proteins and other molecules. According to a complexation scenario proposed earlier, a steady-state ensemble of finite-sized droplets of a metastable solute-rich liquid can emerge in a solution, if the solute molecules can form transient complexes with each other and/or solute. Here we solve for the thermodynamics of an explicit model of a self-associating fluid in which particles can form transient dimers. We determine ranges of parameters where two distinct dense liquid phases, dimer-rich and monomer-rich respectively, can co-exist with each other and the solute-poor phase. We find that within a certain range of the dimer's binding strength, thermodynamic conditions for mesoscopic clusters to emerge are indeed satisfied. The location and size, respectively, of the corresponding region on the phase diagram are consistent with observation. We predict that the clusters are comprised of a metastable monomer-rich dense liquid, while the bulk solution itself contains substantial amounts of the dimer and exhibits large, pre-critical density fluctuations. Surprisingly, we find that the dense phase commonly observed during the macroscopic liquid-liquid separation should be dimer-rich, another testable prediction. The present findings provide further evidence for the complexation scenario and suggest new experimental ways to test it.

\end{abstract}

\maketitle

\section{\label{intro} Motivation}

Solutions of several proteins are known to host droplets of a protein-rich liquid;~\cite{GlikoJACS2005, Georgalis1999, doi:10.1021/jp068827o, PVL, LLVFPostwaldRipening, YEARLEY20141763, SleutelE546, doi:10.1021/acs.cgd.6b01826, Gliko2005, Yamazaki2154, p53clusters, doi:10.1073/pnas.2202222119} the droplets are often called ``mesoscopic'' clusters in reference to their sub-micron size. The clusters contain a tiny fraction of the protein, but are nonetheless detectable in the bulk using dynamic light scattering (DLS) and, more directly, via oblique microscopy,~\cite{10.1063/1.3592581} owing to their large size. Sightings of individual clusters, by means of advanced microscopies,~\cite{GlikoJACS2005, Yamazaki2154} yield sizes consistent with the distributions of mobilities inferred from bulk measurements. A population of clusters promptly appears in a freshly made solution, {  undergoes coarsening,~\cite{LLVFPostwaldRipening, doi:10.1073/pnas.2202222119}} and remains steady for extended periods of time,~\cite{PAN2007267, LLVFPostwaldRipening} implying the ensemble of the clusters is in equilibrium with the ambient solution. 

Protein solutions commonly exhibit a macroscopic liquid-liquid separation into a solute-rich and solute-poor solution,~\cite{Muschol1997, WoldeFrenkel, PhysRevLett.78.2409} when crystallization is kinetically inhibited. The clusters are, however, found well away from the macroscopic liquid-liquid separation. This raises basic questions as to what sort of liquid the clusters are made of and why they grow so large. Indeed, the clusters' contents must be less stable, in bulk terms, than the surrounding solution; the clusters would otherwise grow macroscopically large. This bulk free-energy cost, in addition to the surface tension, should then keep the cluster size down to several particles at most.~\cite{PVL} In contrast, the clusters measure close to a hundred particle sizes across, within a broad range of solute concentration. Incidentally, that the clusters coexist with the ambient solution within a substantial density range excludes a possibility that the clusters could be micelle-like objects. {  For example, Pan et al.~\cite{PVL} report clusters within the range 150-334 mg/ml of the concentration of the solute in the bulk solution; the actual range may be even greater. In contrast, the concentration of a micelle-hosting solution cannot be increased much beyond the critical micelle concentration, at which micelles just begin to appear.}

The microscopic origin of the clusters remains an open question, despite their significance in pathological conditions~\cite{Yange2015618118, https://doi.org/10.1002/aic.14800, UZUNOVA20101976} and their apparent ubiquity: The mesoscopic clusters have now been observed not only for proteins, but also for non-protein solutes,~\cite{doi:10.1021/acs.cgd.7b01299, doi:10.1021/acs.chemmater.5c00679} as well as in non-aqueous solutions.~\cite{doi:10.1021/acs.cgd.5c00370} More recently, the formation mechanism of clusters has been mentioned~\cite{CL_NatureComm} in the broader context of biomolecular condensates, as underlying membrane-less organelles~\cite{Shineaaf4382} or even the emergence of early life. Indeed, condensates break  spatial symmetry by providing locally a distinct chemical environment~\cite{Dai2024, doi:10.1021/acs.accounts.4c00114, doi:10.1021/acs.chemrev.4c00138, doi:10.1021/jacs.4c08946}  even though there may be no physical barrier separating this environment from the surrounding liquid, as in the form of a membrane made of a separate molecular species. 

Pan et al.~\cite{PVL} put forth a microscopic scenario in which the solutes can form transient dimers that live long enough to kinetically stabilize a sizable inclusion of a dimer-rich solution. The characteristic size of the inclusion, according to the solution of a simplified reaction-diffusion scheme,~\cite{PVL} is essentially determined by how far a dimer can travel before it decays back into two monomers: $R =(D \tau)^{1/2}$, where $D$ is the diffusivity of the dimer, $\tau$ its lifetime. 

Subsequently, Chan and Lubchenko~\cite{CL_NatureComm} (CL) worked out a complete reaction-diffusion description based on the Ginzburg-Landau-Cahn-Hilliard (GLCH) functional, in which mesoscopic droplets emerge self-consistently. Similarly to Pan et al.~\cite{PVL}, Chan and Lubchenko assume the putative solute-rich phase comprising the clusters is richer in the dimer than the bulk solution, because complexation is generally favored at higher densities. This is not required, however: It is only necessary that the bulk solution and the droplets, respectively, correspond to two distinct phases, generally characterized by  different compositions of the monomer-complex mixture. Since the clusters consist of a distinct phase, they represent a non-perturbative effect. Owing to additional processes in the form of dimer formation and decay, the spatial profiles of the concentrations and chemical potentials of the two species differ radically from what one would expect during conventional nucleation: The chemical potentials of the two species are not spatially uniform even at the critical droplet size $R_0$, which is a stationary configuration! Droplets smaller across than the size $R_0$ decay, while those that are greater continue to capture the solute at arbitrarily large sizes. Since the bulk free energy of the cluster contents is greater than that of the bulk solution, this would seem to conflict with the Second Law. The conflict is, however, resolved by noting~\cite{CL_NatureComm} that the pressure inside a cluster drops as the cluster size is increased. A growing droplet will, thus, eventually implode. It is one of the unique features of this picture that the growth of a phase nucleus is controlled not through the classical mechanism of evaporation,~\cite{Bray, Farkas+1927+236+242, VolmerSchultze+1931+1+22, https://doi.org/10.1002/andp.19354160806, zeldovich1943theory, Frenkel} but, instead, through a mechanical instability.

Direct experimental tests of the complexation scenario from Refs.~\cite{PVL, CL_NatureComm} have been stymied by the smallness of the amount of solute contained within the clusters, about $10^{-5}$ or less, and the difficulty of directly probing their contents. Judging from available experiment, the clusters each contain around $10^5$ particles. A direct simulation, then, might require a system as large as $10^{10}$ particles, and with little opportunity for coarse-graining, it seems. This complication has motivated semi-analytical treatments, including Refs.~\cite{PVL, CL_NatureComm} and the present work.

The continuum treatment due to Chan and Lubchenko (CL),~\cite{CL_NatureComm} though thermodynamically consistent, is not fully constructive in that it does not provide an explicit microscopic model for the liquid mixture. Although coexistences of liquids and/or vapors with distinct compositions are common,~\cite{BRR, silbey2004physical} it is not at all clear whether the phase behaviors of actual mixtures that host mesoscopic clusters are compatible with the setup of the CL study. Indeed, the clusters must be composed of a metastable phase, as already mentioned, but are seen on the super-critical side of the macroscopic liquid-liquid separation, where there should be just one free energy minimum; no other phases seem to be present. Conversely, {\em known} metastability regions are located near the edges of the coexistence region for the macroscopic liquid-liquid separation, far away from where the clusters are seen. 

To address this puzzle, here we test the following microscopic scenario: The clusters consist of a solute-rich liquid that is distinct from the solute-rich liquid observed during the macroscopic liquid-liquid separation. There must be, then, an extra minimum in the free energy that manifests itself through the clusters but may or may not manifest itself as a macroscopic phase. 

In considering the possibility of poly-liquid behaviors in the context of complexation, we are motivated by an earlier study of Talanquer,~\cite{Talanquer2005phase} who showed that associating fluids can exhibit an immensely rich phase behavior, such as a coexistence of three distinct liquid phases. The theory of associating fluids, due to Wertheim and others,~\cite{Wertheim1984fluidsI,Wertheim1984fluidsII, Wertheim1986fluidsIII, Wertheim1986fluidsIV, Chapman1986theory, Joslin1987theory, JCG1988AL, CJG1988AL, CHAPMAN1989} has been used, with considerable success,~\cite{Muller2001saft, FOUAD201662} to quantitatively describe effects of complexation on vapor-liquid and liquid-liquid equilibria. {  Subsequent works have tailored the theory to elucidate effects of binding-induced changes in dispersion interactions between the constituent particles.~\cite{Talanquer2005phase, 10.1063/1.458711, MARSHALL2025114461} A related set of descriptions of biomolecular condensation is embodied in treatments of patchy particles and associative macromolecules.\cite{doi:10.1021/acs.chemrev.2c00814}}

Here we employ a modified version of Talanquer's theory whereby we limit the oligomer size to two particles, while retaining the key feature of the self-associating fluids that the oligomer-to-monomer ratio is determined self-consistently. We readily find a range of the dimer-binding strength, for which an additional free energy minimum appears away from the conventional liquid-liquid separation. This finding, then, supplies a constructive microscopic basis for the semi-empirical methodology from Ref.~\cite{CL_NatureComm}, while providing further support for the complexation scenario. Whenever this additional, metastable minimum corresponds to solute densities that are above the solubility limit of solid aggregates of the solute, it would likely reveal itself only through the clusters. Just this seems to be the case in protein solutions. 

Contrary to expectation,~\cite{PVL, CL_NatureComm} here we find that the metastable solute-rich phase---which would comprise the clusters---is dimer-poor. It is, instead, the ambient solution that has a substantial dimer-to-monomer ratio. The {\em macroscopic} solute-rich liquid is predicted to be dimer-rich. We believe it would be the most immediate test for the present theory to probe for the dimer content of the bulk solution, when the clusters are present, and, separately, for the dimer content of the macroscopic dense phase. Another testable prediction is that in the bulk solution, when the clusters are present, the mole fractions of the monomer and dimer, respectively, should fluctuate strongly. This is because the solution is supercriticial with respect to the macroscopic liquid-liquid separation between the bulk solution and the dimer-rich dense phase. This notion seems to be consistent with DLS data for concentrated lysozyme solutions.~\cite{PVL}

The article is organized as follows: We briefly describe the model in Section~\ref{methods}. We then analyze, in Section~\ref{results}, a small subset of the parameter space that is of direct relevance to the CL scenario. We identify a regime, in which the thermodynamic conditions for the clusters to exist are satisfied and discuss the applicability of these findings to actual solutions. Section~\ref{summary} contains a summary and brief discussion in the broader context of biomolecular condensation.  

\section{Methods}
\label{methods}

The calculation of the free energy and the equation of state closely follows Talanquer,~\cite{Talanquer2005phase} with the modification that a monomer has only one binding site and so there is only one kind of oligomer, the dimer; the calculational details are provided in the Appendix. Here we only list the basic features of the model and introduce key parameters that control the model's phase behavior. 

The elemental particle of the description is called the ``monomer.'' The interaction between two monomers is presented as a sum of three distinct contributions. The first contribution is a steric, hard-sphere like repulsion---denote the corresponding sphere diameter with $\sigma$---and does not depend on whether the monomers are standalone or bound within oligomers. The second contribution embodies a short-range attraction through a van der Waals-like potential of mean force:
\begin{equation}
	\phi_{ij}(r) =
	\begin{cases}	
		0,  & r \leqslant \sigma\\
		-4\epsilon_{ij} \left( \dfrac{\sigma}{r} \right)^6,  & r > \sigma
\end{cases}
\label{eq:PotentialAttraction}
\end{equation}
where the subscripts $i$ and $j$ each can stand either for ``$m$'' (``monomer'') or ``$d$'' (``dimer''). By construction, \(\phi_{{mm}}\) corresponds to the interaction between two standalone monomers. The potential \(\phi_{{md}} \equiv \phi_{{dm}} \) quantifies the interaction between a free monomer and a monomer that is bound within a dimer. The potential \(\phi_{{dd}}\) corresponds to the interaction between two monomers that are each part of a respective dimer. One may further define the following parameters 
\begin{equation} \label{Dmd} \Delta_{md} = \epsilon_{mm} + \epsilon_{dd} - 2\epsilon_{md}
\end{equation}
and \begin{equation} \label{dmd} \delta_{md} = \epsilon_{md} - \epsilon_{dd}.
\end{equation}
The quantity \(\Delta_{md}\) can be thought of as a de-mixing energy for monomers and dimers, while the parameter \(\delta_{md}\) allows one to parameterize the energy change, if any, arising when a dimer makes a contact with a monomer instead of another dimer.

By construction, each monomer has one attractive site that can be thought of as a ``sticky patch.'' Two monomers can transiently bind to form a dimer, when their respective patches face one another. The potential of mean force for the monomer-monomer binding is a short range attractive potential approximated with a well of depth $\epsilon_\text{bond}$.~\cite{JCG1988AL, Talanquer2005phase}

We intend the present work for liquid solutions. Still, when a solute-poor phase and a solute-rich phase co-exist, we will often refer to the former and latter as the ``vapor'' and ``liquid,'' respectively, to simplify the prose.  Hereby, we regard the buffer as a dielectric continuum, possibly hosting mobile ions. In the same spirit, we will refer to the osmotic pressure of the solute as simply ``the pressure.'' {  To avoid ambiguity, we note that the term “coexistence” can have two meanings that are closely related but not equivalent. “Coexistence” is often used to delineate regions on the phase diagram where there is a macroscopic phase separation between two mutually-stable phases. The region itself can be thought of as the totality of tie-lines that correspond to distinct mole fractions of the two phases. More generally, “coexistence” refers to a simultaneous existence of two or more phases that may or may not be equally stable. Formally, there is a separate phase per each distinct convex-down portion of the free energy, the distinction defined by an instability or an outright gap in the density of states. Any system exhibiting two or more such portions represents a phase coexistence and ergodicity breaking.~\cite{HLKnowledge} Even if two phases are not equally stable, they can still be observed to coexist transiently, as, for instance, when one adds ice to water prepared above melting temperature. (The temperature and/or other parameters may not be uniform since the system is away from equilibrium.)  Only when the much stronger condition of mutual equilibrium is satisfied, the co-existence can last for long times, subject to various coarsening phenomena.} 

{\em Units}. The magnitude of monomer-monomer interaction (\(\epsilon_{mm}\)) and the diameter of the monomer particle (\(\sigma \)) are adopted as the reference scales for energies and lengths, respectively.  Thus, \(T^*=k_B T/\epsilon_{mm}\), \(\Delta_{md}^*=\Delta_{md}/\epsilon_{mm}\), \(\delta_{md}^*=\delta_{md}/\epsilon_{mm}\), \(\epsilon_\text{bond}^*=\epsilon_\text{bond}/\epsilon_{mm}\). The total number density $\rho$ of the monomer, both standalone and bound, becomes \(\rho^*=\rho\sigma^3\). Note the three dimensionless quantities $\Delta_{md}^*$, $\delta_{md}^*$, and $\epsilon_\text{bond}^*$ fully specify the energetics of the model.

\section{Results and discussion}
\label{results}

\begin{figure}[!tb]
\centering
\includegraphics[width=\columnwidth]{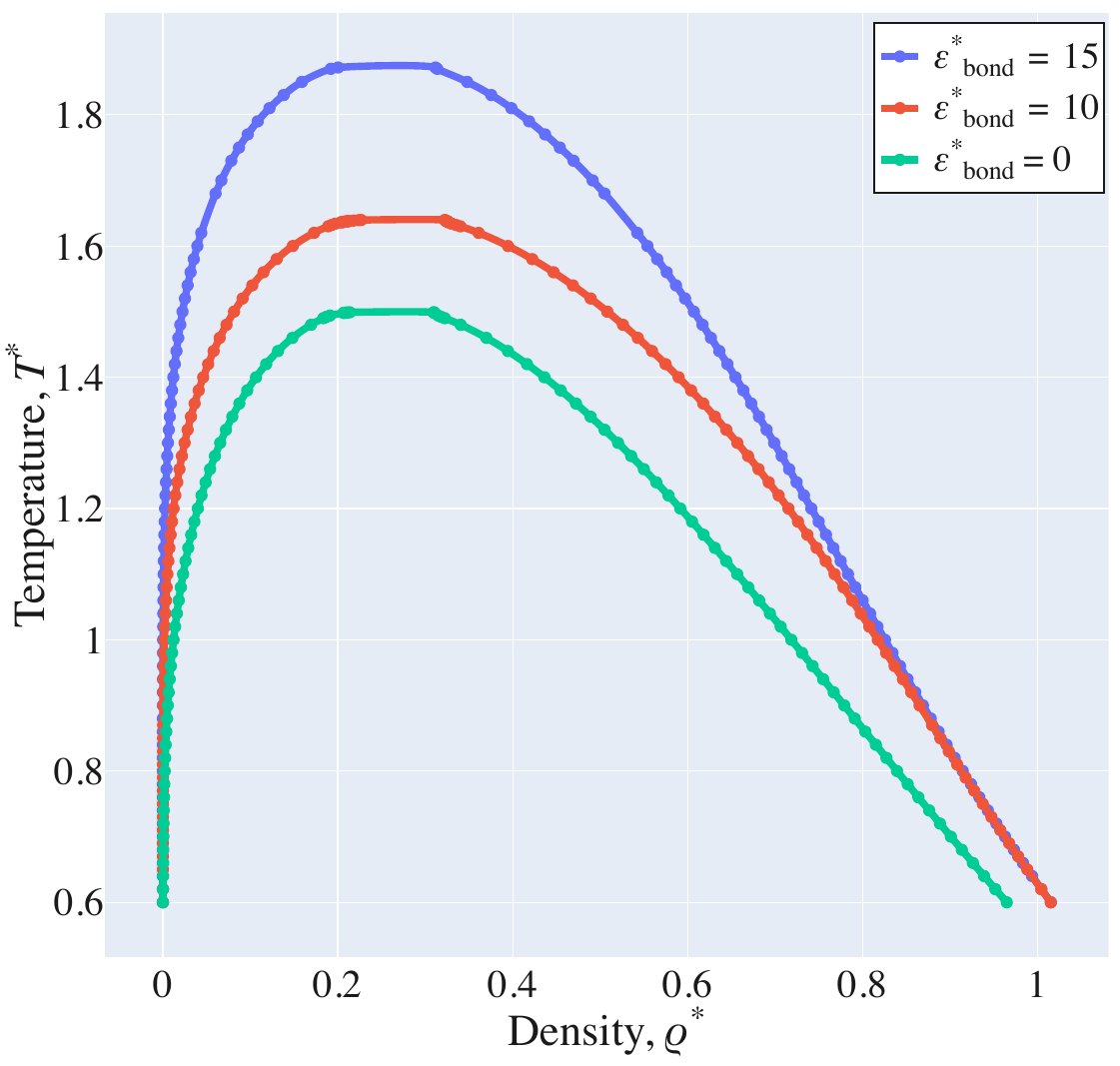}
\caption{Phase diagrams for the simple case 
  \(\Delta^*_{md}=\delta^*_{md}=0\) and select values of
the binding energy  \(\epsilon^*_\text{bond}\).}
\label{fig:DiagramDescription}
\end{figure}

We start off by showing, in Fig.~\ref{fig:DiagramDescription}, the phase diagram for the simple case \(\epsilon^*_{mm} = \epsilon^*_{md} = \epsilon^*_{dd} = 1\)---which implies \(\Delta^*_{md} = \delta^*_{md} = 0\)---and a small set of select values of the binding energy $\epsilon_\text{bond}^*$. This system should exhibit a phase coexistence between a vapor-like and liquid-like state, respectively, as would, for example, the van der Waals gas.  As in Refs.~\cite{Talanquer2005phase} or \cite{CL_LG}, {  we detect distinct phases by finding convex-down portions in the grand-canonical free energy}, per unit volume: 
\begin{equation} \label{negp}
    \Omega/V = f(\rho, T) - \mu \rho = -p,
\end{equation}
as a function of the total density $\rho$, at fixed temperature $T$. Here, $f$ is the spatial density of the Helmholtz energy and $\mu$ is the externally imposed chemical potential. {  In the usual fashion,~\cite{Talanquer2005phase, CL_LG, HLKnowledge, LSurvey} any two phases can be brought to mutual equilibrium, if one can draw a common tangent to the two respective convex-down portions. The construction itself is not affected by the value of $\mu$, which, nonetheless, can be varied to tune the stability of a particular phase.} 

We indicate in the second equality in Eq.~\eqref{negp} that the grand canonical potential per unit volume is also equal to the negative pressure,~\cite{LLstat} since $\Omega \equiv E - TS - \mu N = - p V$. The depth of a minimum, then, corresponds to the equilibrium value of the negative pressure in the respective phase, at the given values of temperature and the chemical potential. Any two minima of equal depth correspond to two phases in mutual equilibrium.

Consistent with expectation, the simple case \(\Delta^*_{md} = \delta^*_{md} = 0\) yields a coexistence region that has a characteristic dome-like shape. When binding is added on top of an already existing isotropic attraction, the phase coexistence region moves up to higher temperatures. In other words, the introduction of association into the system uniformly over free and bound monomers, respectively, effectively enhances the overall attraction. Although some dimer is present, the mole fraction $X$ of the unbound monomer is numerically close to 1 both in the dilute and dense phase, consistent with Ref.~\cite{Talanquer2005phase} 

If one includes effects of binding on the inter-mononer interactions, however, the behavior of a self-associating fluid can become staggeringly complex.~\cite{Talanquer2005phase}  Here we focus on three sets of parameters that are of direct relevance to the complexation scenario from Ref.~\cite{CL_NatureComm} Specifically, we ask whether there is a combination of parameters for which: 

(a) Two distinct dense-liquid phases are possible, and 

(b) One of these dense-liquid phases is metastable well outside the region where the dilute phase and the other dense phase coexist at ambient conditions. 

\begin{figure}[!tb]
    \centering
    \includegraphics[width=\columnwidth]{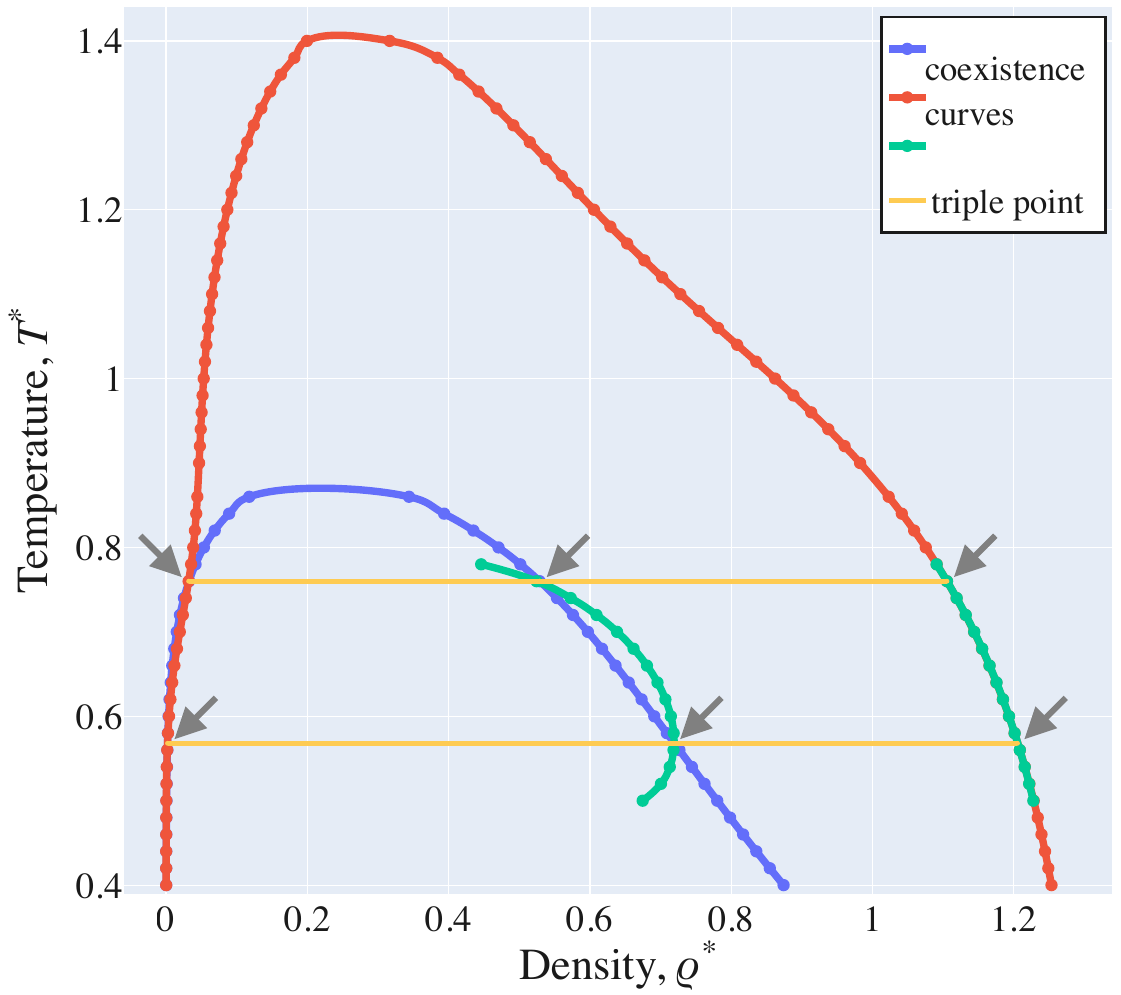}
    \caption{{  An instance of a phase diagram with three distinct phase equilibria indicated, respectively, by the red dome, blue dome, and a set of two green lines. A segment of the blue dome on the left is obstructed by the red dome, while a segment of the red dome on the right is obstructed by the one of the green lines. The system has two triple points that are located, respectively, at the two yellow tie-lines, pertinent concentrations indicated by two sets of three arrows.  See text for an explanation of the three phase equilibria. } \(\Delta^*_{md} = 0.4\), \(\delta^*_{md} = 0.05\) and \(\epsilon^*_\text{bond} = 9.25\).  }
    \label{fig:ExDiagTriple}
\end{figure}

\begin{figure}[!tb]
    \centering
    \includegraphics[width=0.7\columnwidth]{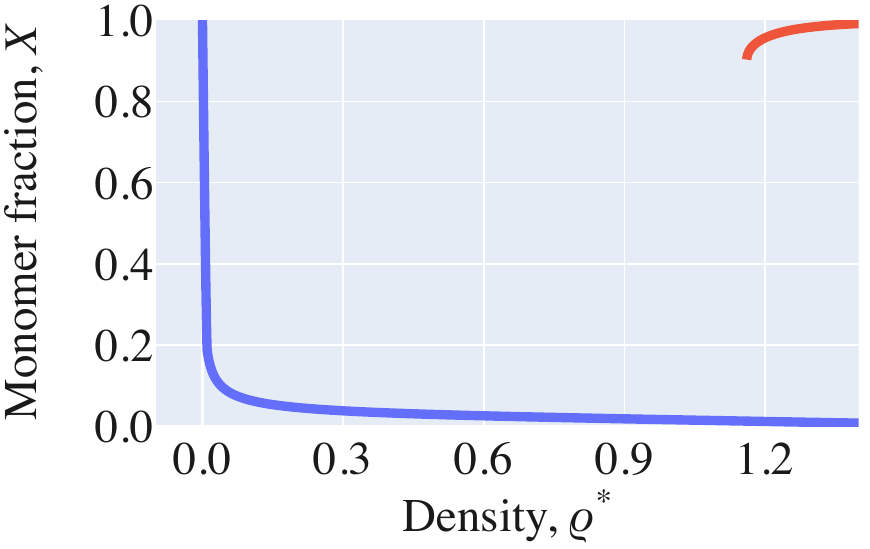}
    \caption{Fraction of the monomer as a function of the total density, at \(T^*=0.64\), for the phase diagram in Fig.~\ref{fig:ExDiagTriple}. \(\Delta^*_{md} = 0.4\), \(\delta^*_{md} = 0.05\) and \(\epsilon^*_\text{bond} = 9.25\) }
    \label{XrhoT064}
\end{figure}

\begin{figure}[!tb]
    \centering
    \includegraphics[width=0.7\columnwidth]{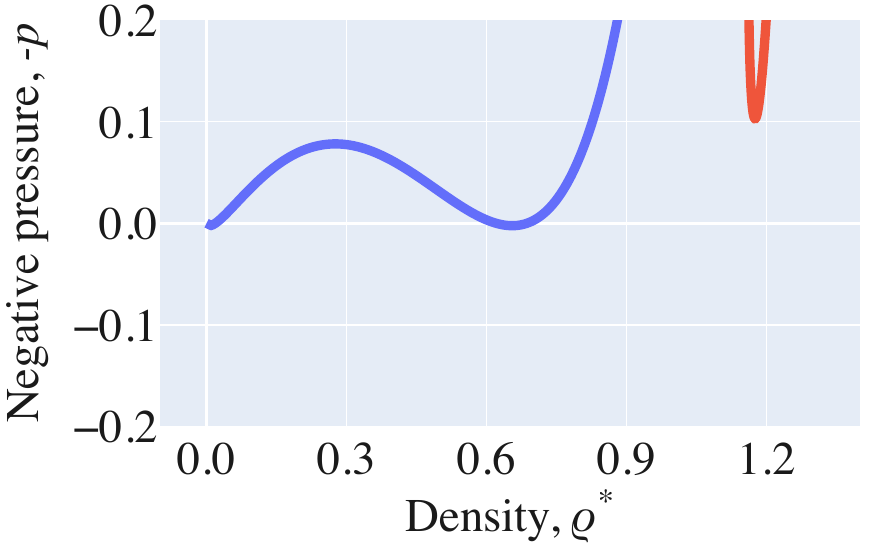}
    \caption{The negative pressure from Eq.~\eqref{negp} as a function of the total density at \(T^*=0.64\). The conditions are the same as in Fig.~\ref{fig:ExDiagTriple}: \(\Delta^*_{md} = 0.4\), \(\delta^*_{md} = 0.05\) and \(\epsilon^*_\text{bond} = 9.25\). The chemical potential $\mu$ is set so that the vapor and the dimer-rich liquid are in mutual equilibrium. Note that at this temperature, each pair of the three phases can be mutually equilibrated. }
    \label{frhoT064}
\end{figure}

Consonant with Talanquer's results,~\cite{Talanquer2005phase} the present model indeed can exhibit two distinct dense-liquid phases, {  in addition to the dilute, non-condensed phase}. As our first illustration, we employ the set of parameters that yield the phase diagram shown in Fig.~\ref{fig:ExDiagTriple}. 
To elucidate the compositions of the two dense phases, we display the density dependence of the monomer fraction in Fig.~\ref{XrhoT064} at the $T^* = 0.64$ tie-line, which is an intermediate temperature between the two tie-lines shown in yellow in Fig.~\ref{fig:ExDiagTriple}. Note the $T^* = 0.64$ tie-line crosses through all three phases. The two branches in Fig.~\ref{XrhoT064}, color-coded blue and red, respectively, correspond to two distinct branches of the free energy density, which are shown in Fig.~\ref{frhoT064} using the same color scheme. In the latter Figure, we chose a particular value of the chemical potential $\mu$, at which the dilute ($\rho^*$ near zero) and intermediate-density ($\rho^* \simeq 0.7$) phases, respectively, are in mutual equilibrium. According to Fig.~\ref{XrhoT064}, the dilute phase is monomer rich, while the intermediate-density phase is, in contrast, dimer-rich. The densest phase ($\rho^* \simeq 1.2$), on the other hand, is monomer rich.  

{  For the reader's reference, we provide a brief description of how the phase coexistences and/or mutual equilibria illustrated in Fig.~\ref{fig:ExDiagTriple} come about. To simplify the prose, let us label the three phases corresponding to the three free-energy minima in Fig.~\ref{frhoT064} as 1, 2, and 3, counting left to right. There are three possible two-phase equilibria at temperatures corresponding to Fig.~\ref{frhoT064}: 1-2, 2-3, and 1-3. The red dome in Fig.~\ref{fig:ExDiagTriple} corresponds to the 1-3 equilibrium. The blue dome corresponds to the 1-2 equilibrium, and the set of two green lines corresponds to the 2-3 equilibrium. A portion of the blue dome on the left is obstructed by the red line and a portion of the red dome on the right is obstructed by one of the green lines.

At temperatures above the top yellow line in Fig.~\ref{fig:ExDiagTriple}, minimum 2 is above the double tangent for phases 1 and 3. (At temperatures above the top of the blue dome there is no phase 2 in the first place.) Consequently, the 1-3 equilibrium is stable above the top yellow line, while the 1-2 and 2-3 equilibria, if any, are metastable because phase 2 itself is never stable at these temperatures. Now, as the temperature is lowered to the value given by the top yellow line, minimum 2 moves down just to the extent that one can now draw a common tangent to all three minima at the same time, implying the chemical potential can be tuned to bring all three phases to mutual equilibrium at the same time. Thus the top yellow line corresponds to a triple point. Lowering temperature below this triple point will move minimum 2 below the double tangent for phases 1 and 3. Consequently, the 1-3 equilibrium becomes metastable because phase 2 is the lowest in free energy whenever phases 1 and 3 are in mutual equilibrium. Conversely, the 1-2 and 2-3 equilibria are now stable. In a non-trivial turn of events, further lowering temperature eventually causes minimum 2 to move up and find itself again on the common tangent for minima 1 and 3, thus resulting in another triple point, which is indicated by the bottom yellow line. Below this triple point, the 1-3 equilibrium becomes stable again.}   

From here on, we will refer to the two dense phases as ``dimer-rich'' and ``monomer-rich.'' The latter is denser than the former, both in absolute terms and in terms of the total concentration of the distinct species. The red dome in Fig.~\ref{fig:ExDiagTriple} delineates the phase equilibrium between the vapor and the monomer-rich liquid, respectively. The blue dome, on the other hand, corresponds to the mutual equilibrium of the vapor and the dimer-rich liquid. These two equilibria each end in a critical point. Thus the respective vapor-liquid transitions can each be effected in either a continuous or discontinuous fashion. On the other hand, the transition between the two dense phases is always discontinuous, since the low-density and high-density portions of the pertinent coexistence region never cross, consistent with the free energy having two separate branches for the two phases. 

We note in passing that instances of three-liquid coexistences in mixtures have been reported~\cite{doi:10.1073/pnas.2102512118} but do not necessarily involve complexation.

\begin{figure}[!tb]
    \centering
    \includegraphics[width=\columnwidth]{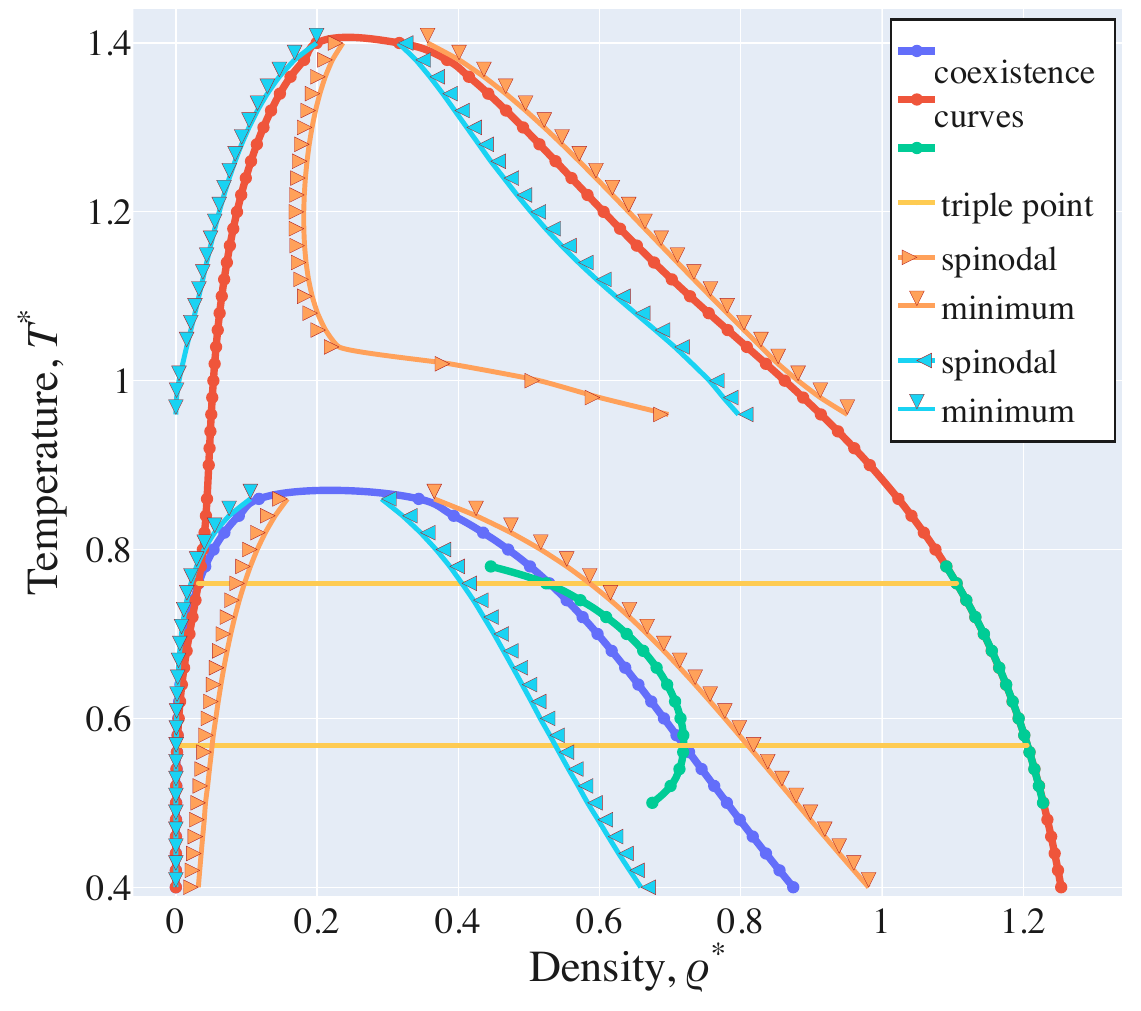} 
    \caption{Phase diagram from Fig.~\ref{fig:ExDiagTriple}, supplemented by the stability limits and the corresponding density values of the respective stable phase. \(\Delta_{md}=0.4\),
      \(\delta_{md}=0.05\) and \(\epsilon_\text{bond}=9.25\)}
    \label{fig:ExDiagComplete}
\end{figure}

Though complicated, the phase diagram in Fig.~\ref{fig:ExDiagTriple} is relatively surveyable. Thus we will employ this particular parameter set to explain the procedure used here to identify candidate regions on the phase diagram where thermodynamic conditions for the clusters to exist are met. The result of this procedure is shown in Fig.~\ref{fig:ExDiagComplete}. For reasons that will become clear shortly, we will primarily focus on the high-temperature portion of the phase diagram, $T^* \gtrsim 1.0$. Compared with Fig.~\ref{fig:ExDiagTriple}, we have added sets of lines color-coded light-blue and orange, respectively; triangular symbols are placed along these added lines to help distinguish them from the equilibrium phase boundaries. The orange and light-blue lines inside the red dome show the mechanical stability limits of the vapor and the monomer-rich liquid, respectively. In the usual way,~\cite{HLKnowledge} these limits are determined by the location of the respective spinodals, $\partial^2 (-p)/\partial \rho^2 = 0$. The location of each spinodal yields the value of the density of the pertinent marginally-stable phase itself. For instance, the light-blue line inside the red dome, at $T^* \gtrsim 1.0$, gives the density of the monomer-rich liquid when the latter liquid is marginally stable.

{  The phrase “marginally stable phase” means that there exists a free energy minimum defining this phase but the free energy barrier to escape the minimum is vanishingly small. One way to think how the marginal stability comes about here is this: Let us start out, on the phase diagram in Fig.~\ref{fig:ExDiagComplete}, with a pure dilute phase on the l.h.s. of the red dome. Set temperature at $T^* = 1.1$, for the sake of concreteness. At these values of temperature and chemical potential, respectively, the monomer-rich liquid would be as stable as the dilute phase. Next we move straight to the left on the phase diagram by removing the solute. Equivalently, this is realized by decreasing the chemical potential at constant temperature. In the process, the free-energy minimum corresponding to the monomer-rich liquid becomes destabilized relative to the dilute phase, while becoming more shallow. Mathematically, this comes about since we are adding a straight line with a positive slope to a bi-stable potential. Eventually, the r.h.s. minimum defining the monomer-rich phase becomes marginally stable. At this moment, we place two light-blue points on the phase diagram: One point is placed at the concentration of the bulk solution and $T^* = 1.1$, the other point at the concentration of the (marginally stable) monomer-rich phase and $T^* = 1.1$. The construct is performed likewise at other temperatures, to yield the two light-blue lines in Fig.~\ref{fig:ExDiagComplete} (at $T^* \gtrsim 0.95$). Moving to the left of the l.h.s. light-blue line will remove the minimum corresponding to the monomer-rich phase altogether; the respective portion of the free energy will now be a convex-down, positively sloped curve. }

To summarize, the light-blue line on the outside and left from the red dome, at $T^* \gtrsim 1.0$, gives the value of the density of the {\em vapor}, for which the monomer-rich liquid is marginally stable. Thus, the very narrow region between this light-blue line and the adjacent portion of the red dome is special in the present context: The solution within this region meets both thermodynamic conditions (a) and (b), stated above, for the clusters to exist. For completeness, we have computed the rest of the boundaries where metastable phases coexist with respective stable phases; {  in each pair, the metastable phase is the less stable phase by construction}. The color-coding in Fig.~\ref{fig:ExDiagComplete} should be self-explanatory.

\begin{figure}[!tb]
    \centering
    \includegraphics[width=\columnwidth]{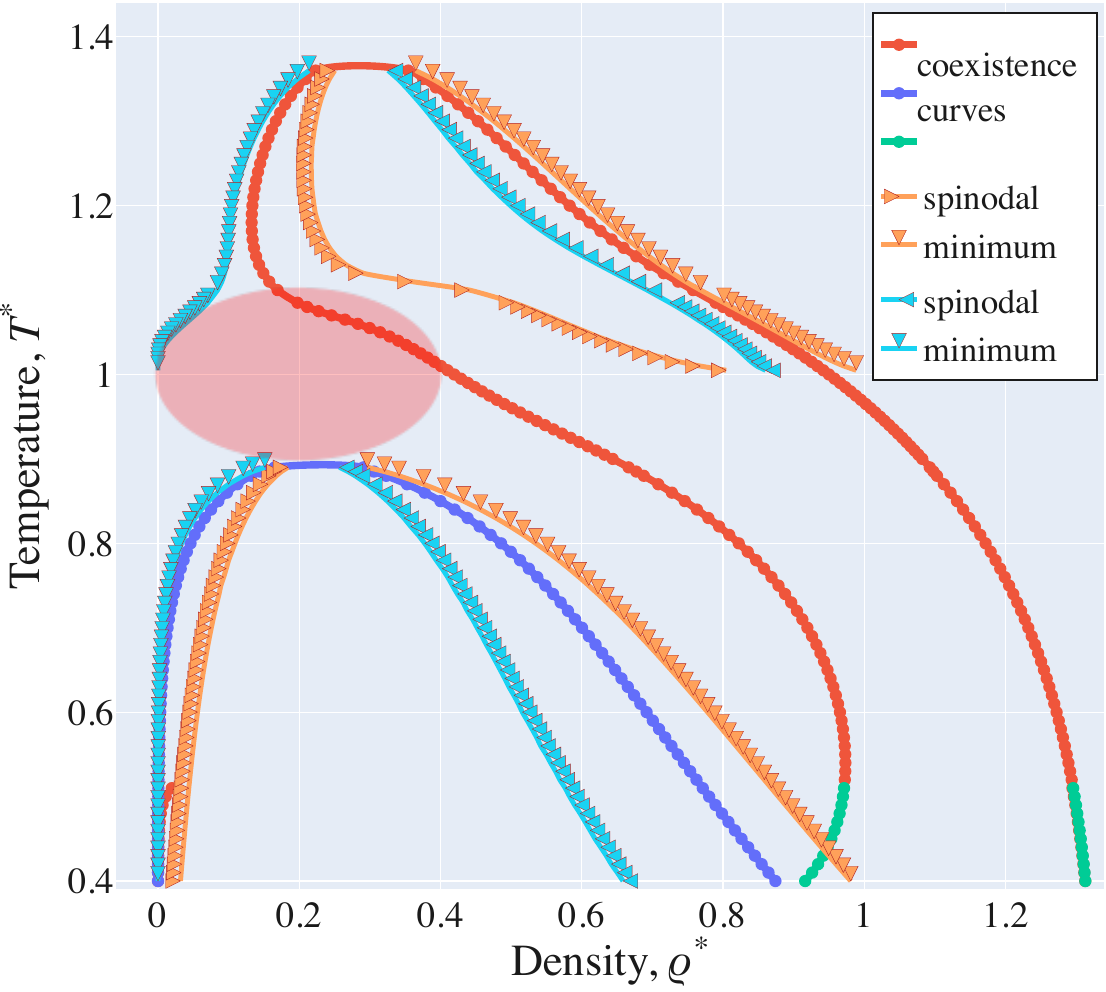}
    \caption{Phase diagram for the parameter set \(\Delta_{md}=0.4\),
      \(\delta_{md}=0.05\) and \(\epsilon_\text{bond}=9.60\). The shaded oval-shaped region approximately indicates the region where mesoscopic clusters would be thermodynamically allowed. {  See the text for the explanation of the equilibria.}}
    \label{fig:Diag96}
\end{figure}

The set of conditions in Fig.~\ref{fig:ExDiagComplete} where clusters could be potentially found corresponds to a very narrow density interval that would seem to be unlikely to be detected in experiment. Shape-wise, this region is not at all like ones seen in experiment to date. In an attempt to broaden the region, we next increase the dimer's binding energy $\epsilon^*_\text{bond}$, so as to shift the equilibrium toward higher densities. According to Fig.~\ref{fig:Diag96}, already increasing the value of $\epsilon^*_\text{bond}$ from 9.25 to 9.6 suffices to move the coexistence between the bulk-solution and the monomer-rich dense liquid to significantly higher densities, within a substantial temperature range $1.0 \lesssim T^* \lesssim 1.1$, c.f. Fig.~5 of Ref.~\cite{Talanquer2005phase} At the same time, the relative stability of the bulk solution relative to the {\em marginally}-stable monomer-rich liquid is affected only weakly, apparently. This results in a dramatically increased size of the region where clusters could be present; the region itself is roughly indicated by the shaded ellipse in Fig.~\ref{fig:Diag96}. This finding is the main result of this work.

\begin{figure}[!tb]
    \centering
    \includegraphics[width=0.7\columnwidth]{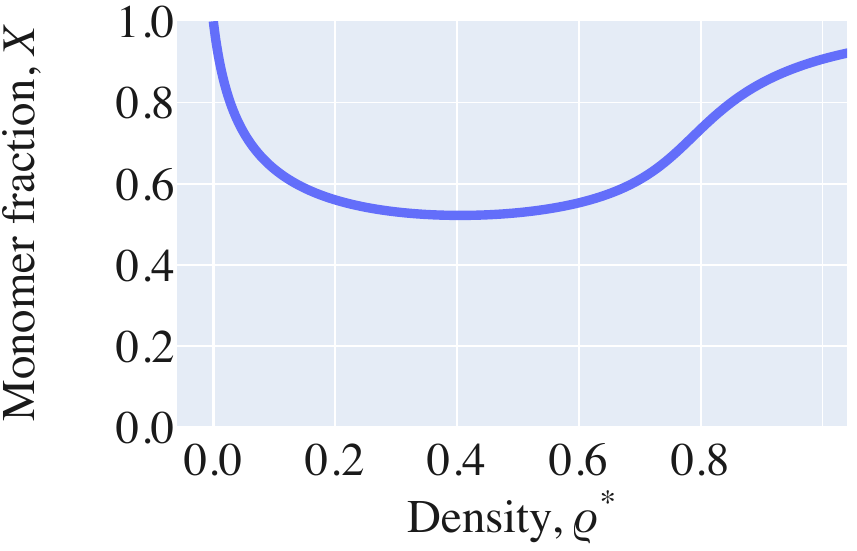}
    \caption{Fraction of the monomer as a function of the total density, for the phase diagram in Fig.~\ref{fig:ExDiagComplete}, at \(T^*=1.04\). \(\Delta^*_{md} = 0.4\), \(\delta^*_{md} = 0.05\) and \(\epsilon^*_\text{bond} = 9.60\). }
    \label{Xrho96}
\end{figure}

\begin{figure}[!tb]
    \centering
    \includegraphics[width=0.7\columnwidth]{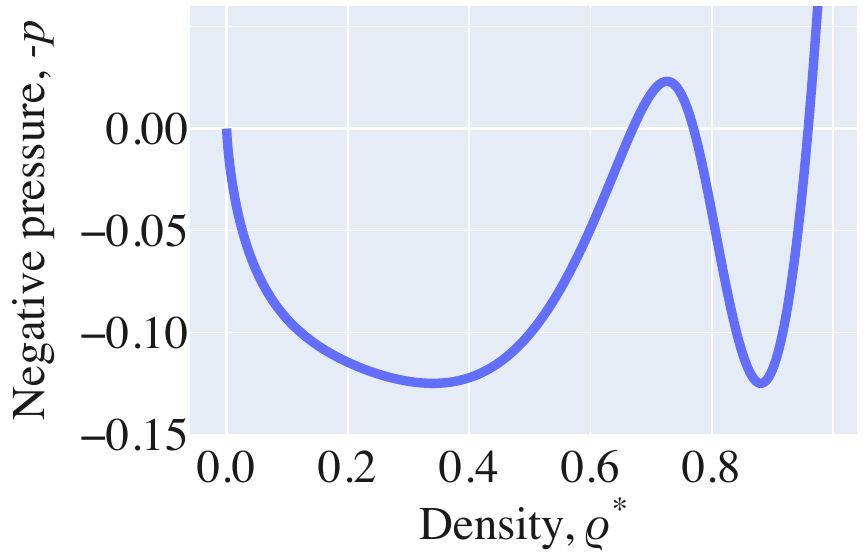}
    \caption{The negative pressure from Eq.~\eqref{negp} as a function of total density, for the phase diagram in Fig.~\ref{fig:ExDiagComplete}, at \(T^*=1.04 \). \(\Delta^*_{md} = 0.4\), \(\delta^*_{md} = 0.05\) and \(\epsilon^*_\text{bond} = 9.60\). The chemical potential $\mu$ is set so that the bulk solution and the monomer-rich liquid are in mutual equilibrium. }
    \label{frho96}
\end{figure}

Of considerable interest are the composition and the free energy of a cluster-hosting bulk solution. These are illustrated in Figs.~\ref{Xrho96} and \ref{frho96}, respectively. We observe that except at low overall densities, the amounts of monomer and dimer, respectively, in the bulk solution are comparable. We note that the bulk solution remains stable even at relatively large densities, which are close to the critical density for the vapor/dimer-rich liquid equilibrium. Consistent with this, the free energy of the bulk solution at the pertinent densities presents as a rather broad minimum, indicating that the density would be subject to large fluctuations. This finding is consistent with  observation. Indeed, the auto-correlation functions for dynamic light scattering in concentrated lysozyme solutions do exhibit a long, power law-like tail.~\cite{PVL}

{  For the reader's reference, we briefly discuss the phase behavior in the low temperature part of the phase diagram in Fig.~\ref{fig:Diag96}. The blue dome has the same meaning as in Fig.~\ref{fig:ExDiagTriple}, with the difference that this coexistence is now stable throughout. At all temperatures above the upper ends of the green lines, the dilute phase is at best marginally stable with respect to the monomer-rich dense liquid. Thus at these temperatures, only the dimer-rich dense liquid can be in equilibrium with the monomer-rich dense liquid. At temperatures corresponding to the green lines, the dilute phase becomes metastable with respect to the monomer-rich liquid, while the two phases can be brought to a metastable equilibrium with each other. The low density side of the respective co-existence region is indicated by the partially obstructed red line in the left bottom corner. The corresponding high density side of the coexistence is obstructed by the r.h.s. green line. The two green lines indicate the (stable) equilibrium between, respectively, the dimer-rich and monomer-rich dense liquids.}

We would be remiss not to discuss the putative monomer-rich phase, in the context of actual systems. Within the temperature range $1.0 \lesssim T^* \lesssim 1.1$, where clusters are thermodynamically allowed, the density of a stable monomer-rich liquid would vary within the range $0.8 \lesssim \rho^* \lesssim 1.0$, according to Fig.~\ref{fig:ExDiagComplete}. At such high densities, the monomer-rich liquid would be close in stability to the crystal, see for instance Ref.~\cite{L_AP}. (The crystal state was not considered in this work.) This notion suggests that such a monomer-rich liquid would be difficult to observe as a macroscopic phase. The notion is also consistent with the clusters, when present, being preferred nucleation sites for crystallization.~\cite{C4FD00217B, Yamazaki2154, doi:10.1021/acs.cgd.6b01826} What about higher temperatures, $T^* \gtrsim 1.0$? At these higher temperatures, the monomer-rich liquid may well be more stable than the corresponding crystal and, as such, it could become observable. While this possibility is hard to exclude generally, it seems unlikely to materialize in protein solutions. Indeed, in absolute terms, the range $1.0 \lesssim T^* \lesssim 1.1$ corresponds to 20-to-30 degrees, the upper edge of the range being easily $40^\circ$ above room temperature. At such conditions, most proteins would denature, which would likely lead to a host of additional aggregation processes. 

We note the positive signs and magnitudes of the quantities, $\Delta_{md} > \delta_{md} > 0$ we have employed here are the likely ones to be realized in actual systems.~\cite{Talanquer2005phase} Indeed, the definitions \eqref{Dmd} and \eqref{dmd} imply 
\begin{equation}
    \Delta_{md} = \epsilon_{mm} - \epsilon_{md} -\delta_{md}.
\end{equation}
Thus, the inequalities $\Delta_{md}> \delta_{md} > 0$ are equivalent to saying that $\epsilon_{mm} > \epsilon_{md} > \epsilon_{dd}$. That is, an individual monomer interacts progressively more weakly with other particles as the monomer becomes bound within a dimer. This would be consistent with the individual monomer becoming progressively less polarizable as its electronic configuration becomes stabilized following binding.   

\begin{figure}[!tb]
    \centering
    \includegraphics[width=\columnwidth]{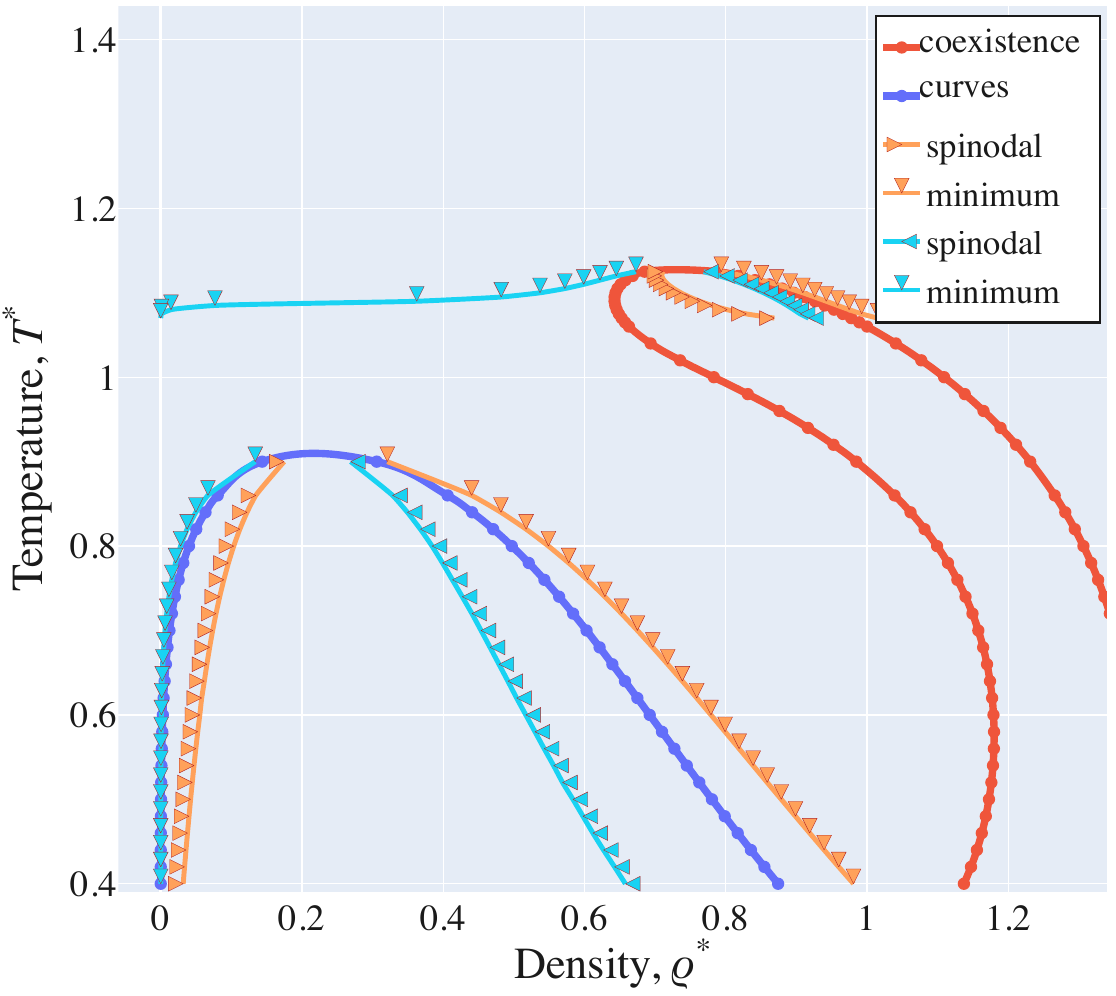}
    \caption{Phase diagram for \(\Delta_{md}=0.4\),
      \(\delta_{md}=0.05\) and \(\epsilon_\text{bond}=10.00\)}
    \label{fig:Diag100}
\end{figure}

Now, increasing the binding energy even further will broaden the density range where the clusters are thermodynamically allowed, while shrinking the corresponding temperature range, as we illustrate in Fig.~\ref{fig:Diag100}. Thus we conclude that in order to exist, clusters may require some fine-tuning of intermolecular interactions, at least for the very simplified types of force fields we have considered here. This said, many molecules exhibit a larger set of interactions than we considered here. This may lead to substantially broader ranges of thermodynamic conditions at which clusters would be allowed. 

\section{Concluding Remarks}
\label{summary} 

We have confirmed, on a constructive microscopic basis, the thermodynamic consistency of the Chan-Lubchenko scenario~\cite{CL_NatureComm} for the formation of the mesoscopic clusters. The origin of these liquid-like aggregates remains poorly understood, while experimental probes and brute force simulations are challenging. We have adapted Talanquer's version~\cite{Talanquer2005phase} of the statistical associating fluid theory~\cite{Wertheim1984fluidsI,Wertheim1984fluidsII, Wertheim1986fluidsIII, Wertheim1986fluidsIV, Chapman1986theory, Joslin1987theory, JCG1988AL, CJG1988AL, CHAPMAN1989, Muller2001saft, FOUAD201662} to a case where the oligomer size is limited to two monomers. We have identified a relatively narrow parameter range where the requisite thermodynamic conditions for the clusters to be present are satisfied: On the one hand, the solution exhibits a solute-rich metastable phase, which would constitute the clusters. On the other hand, the location of the corresponding region on the phase diagram of the solution is consistent with observation. We predict that the solute-rich contents of the clusters are distinct from the macroscopic solute-rich phase seen during the liquid-liquid separation. Our analysis further indicates that the dense liquid constituting the clusters should be rich in the monomer itself, while the bulk solution should contain substantial amounts of the oligomer. The observed, macroscopic protein-rich liquid is predicted to be quite rich in the oligomer. In view of difficulties in directly probing the contents of the clusters, it appears that testing for the oligomer content of the solution---when the clusters are present---and, separately, for the oligomer content of the macroscopic dense liquid would be good starting points for testing the present scenario.  

{  It seems instructive to look at the present results from the perspective of the Gibbs phase rule.~\cite{ma17246048} The presence of more than one species does not, in itself, amount to an additional degree of freedom, since the composition in each phase is strictly tied to the temperature and chemical potential. It is not at all obvious that the present system should exhibit three distinct minima in its free energy such that each minimum corresponds to a fluid-like phase. Incidentally, a mechanism for emergence of multiple liquid phases for mixtures of particles with complicated shapes has been reported.~\cite{doi:10.1021/acs.jpclett.2c03138} Those phases differ by the extent to which symmetry is lowered locally, even though the translational symmetry is preserved, implying the solution is still a fluid. Likewise, Talanquer's analysis and the present results alike indicate that the symmetry breaking caused by the effects of binding on dispersion interaction is sufficient to give rise to an additional free energy minimum.}

We reiterate that the mesoscopic clusters represent a special route toward biomolecular condensation. Indeed, the resulting droplets of the condensate would not require additional regulation, on the cell's part, to prevent unlimited growth, in contrast with the macroscopic dense liquid. This would seem to make clusters a relatively robust way to create locally chemically-heterogeneous environments, possibly indicating their significance for early life forms. {  The present setup arguably represents the simplest arena for studying mesoscopic aggregation. In actual settings, mesoscopic aggregates may exhibit internal structure~\cite{Wu2025} and appear to be part of a broad continuum of aggregates that could form in response to various stimuli.~\cite{Lan2023, Kar2024, Yanas2024, HOFFMANN2025168987}}

We have seen that the thermodynamic conditions, for the clusters to be present, are met within a rather narrow range of the inter-particle interactions. In addition, Chan and Lubchenko~\cite{CL_NatureComm} treat the parameters of the Landau-Ginzburg free energy, on the one hand, and the kinetic coefficients, on the other hand, as independent. But the respective sets of parameters should both derive from the same molecular forces, which amounts to a constraint. This additional constraint, though not rigid, should further narrow the set of systems that could exhibit the clusters.

Though alarming at a first glance, the notion of the constraint is actually consistent with the fact that most sightings of the clusters have been in protein solutions, so far. Evolutionary selection criteria for biological molecules do include properties with respect to aggregation~\cite{10.1042/BST20120160, https://doi.org/10.1038/sj.embor.7401034, doi:10.1126/science.1123539, https://doi.org/10.1002/aic.14800} and, specifically, biocondensation~\cite{KeyportKik2024, Patil2023, 10.1371/journal.pcbi.1012826} since the cytoplasm is a very crowded milieu. Multiple examples of interplay between condensation and the formation of solid aggregates have been reported. For instance, mesoscopic clusters, when present, appear to be preferred nucleation centers for solid-type protein aggregates, such as crystals or fibers, among other things.~\cite{C4FD00217B, Yamazaki2154, Yange2015618118, p53clusters, C2FD20058A, doi:10.1021/acs.cgd.6b01826} Indeed, the condensation may help lower the barrier for nucleating denser yet aggregates.~\cite{doi:10.1063/1.1943389}  Conversely, biomolecular condensation can compete with the formation of solid aggregates that nucleate via alternative mechanisms.~\cite{DAS2025} In any event, the significance of the clusters and of their formation mechanism extend beyond biological systems. The clusters are an entirely general, yet surprising, phase-ordering behavior that has upended classical notions of heterophase fluctuations~\cite{Frenkel, Bray} and phase nucleation.~\cite{Farkas+1927+236+242, VolmerSchultze+1931+1+22, https://doi.org/10.1002/andp.19354160806, zeldovich1943theory} The present results narrow down the scope of parameters for direct simulational tests of the complexation scenario and the accompanying poly-liquid behaviors; this is a subject for future work. \\

{\bf Acknowledgments}: V.~L. thanks Peter G. Vekilov for many insightful conversations. We gratefully acknowledge the support by the NSF Grants MCB-1518204 and CHE-1956389, the Welch Foundation Grant E-1765, and a grant from the Texas Center for Superconductivity at the University of Houston. We gratefully acknowledge the use of the Maxwell, Opuntia and Sabine Clusters at the University of Houston.  Partial support for this work was provided by resources of the uHPC cluster managed by the University of Houston and acquired through NSF Award Number ACI-1531814.

\appendix

\section{Detailed Description of the Model}

The spatial density of the full Helmholtz free energy is presented as a sum of three terms:~\cite{JCG1988AL, CJG1988AL, Muller2001saft, Talanquer2005phase}
\begin{equation} \label{eq:FreeEnergyFull} f = f_{h} +
  f_{\mathrm{att}} + f_{\mathrm{bond}}.
\end{equation}
These three contributions approximately account for the three contributions to the inter-monomer interaction listed in Section~\ref{methods} of the main text: The \(f_{h}\) term accounts for the harsh repulsion (``$h$'' signifies ``hard sphere''). The term \(f_{\mathrm{att}}\) corresponds to the attractive terms from Eq.~\eqref{eq:PotentialAttraction},  hence the subscript ``att.'' The last term, \(f_{\mathrm{bond}}\), is the contribution due to the association of monomers into complexes.

For the hard sphere term, we use the Carnahan-Starling approximation:~\cite{CarnahanStarling1969}
\begin{equation} \label{eq:FreeEnergyHardSpheres} f_{h} = k_B T\rho\left( \ln(\rho) - 1 + \dfrac{4\eta - 3\eta^2}{(1 - \eta)^2} \right)
\end{equation}
where \(k_B\) is Boltzmann constant, \(T\) the temperature, and \(\rho\) the total number density of the monomer, whether bound or not. The dimensionless quantity \( \eta = \frac{\pi}{6}\sigma^3\rho \) gives the total packing fraction of the solute.

The long-range attractive interactions are approximated by the lowest term in the virial expansion:
\begin{equation} \label{eq:FreeEnergyAttractionRho}
	f_{\mathrm{att}} = -\frac{1}{2} \left( 
	\alpha_{mm}\rho_m^2 
	+ \alpha_{dd}\rho_d^2 
	+ 2\alpha_{md}\rho_m\rho_d
	\right)
\end{equation}
where we present the full number density of the monomer as a sum of two contributions, 
\begin{equation}
   \rho = \rho_m + \rho_d, 
\end{equation}
corresponding to standalone monomers ($\rho_m$) and those bound within dimers ($\rho_d$). Note the number density of the dimer proper is $\rho_2 = \rho_d/2$. The coupling constants reflect that two monomers may interact differently depending on whether any one of them is standalone or, instead, is bound within a dimer:
\begin{equation} \label{eq:AlphaIJ} \alpha_{ij} = - \int d \bm{r}
  \phi_{ij}(r), \qquad i,j=m,d.
\end{equation}
Substituting the potential from Eq.\eqref{eq:PotentialAttraction} into
Eq.\eqref{eq:AlphaIJ} yields 
\begin{equation} \label{eq:AlphaIJsimple}
  \alpha_{ij} = \alpha\epsilon_{ij}\sigma^3     
\end{equation}
where \( \alpha = \frac{16\pi}{3}\).

After we denote the mole fraction of non-bonded monomers with \(X\), the contributions of free and bound monomers, respectively, to the full density can be expressed as:
\begin{equation} \label{eq:Densities}
	\begin{gathered}
		\rho_m = \rho X \\
		\rho_d = \rho \left( 1 - X \right)
	\end{gathered}
\end{equation}
the attractive part of the free energy density can be written as:
\begin{equation}
	f_\text{att} = 
	-\dfrac{\alpha\sigma^3\rho^2}{2} \left( 
	\Delta_{md}X^2 
	+ 2\delta_{md}X 
	+ \epsilon_{dd} 
	\right)
\label{eq:FreeEnergyAttractionX}
\end{equation}
and the quantities  $\Delta_{md}$ and $\delta_{md}$ were defined in Eqs.~(\ref{Dmd}) and (\ref{dmd}), respectively.

The free energy cost $f_{\mathrm{bond}}$ of binding is approximated using the Associating Liquid Theory of
Wertheim, as pertinent to monomers each having a single attractive patch:~\cite{JCG1988AL, CJG1988AL}
\begin{equation}
  f_\text{bond} = k_B T\rho \left( 
    \ln(X) 
    - \dfrac{X}{2} 
    + \dfrac{1}{2} 
  \right)
\label{eq:FreeEnergyBonding}
\end{equation}
Substitution of Eqs.\eqref{eq:FreeEnergyHardSpheres}, \eqref{eq:FreeEnergyAttractionX} and \eqref{eq:FreeEnergyBonding} into Eq.~\eqref{eq:FreeEnergyFull} yields a closed-form expression for the Helmholtz free energy of our self-associating fluid as a function of the total density and the mole fractions of the species.

The mole fraction of the monomer $X$ must satisfy conditions for chemical equilibrium in the system, because the binding is reversible. Specifically, we write the law of mass action in the following form: 
\begin{equation} \dfrac{\rho_2}{\rho_m^2} = \dfrac{\rho_d/2}{\rho_m^2} = K \lambda_\text{att} \label{massact}
\end{equation}
leading, by Eq.\eqref{eq:Densities}, to:
\begin{equation} \label{eq:MassActionEquilibriumX}
\dfrac{1 - X}{\rho X^2} = 2 K \lambda_\text{att}
\end{equation}
The equilibrium constant \(K\) in Eq.~ \eqref{eq:MassActionEquilibriumX} is estimated according to:~\cite{JCG1988AL}
\begin{equation} \label{eq:EquilibriumConstant}
K = 4 \pi g_\text{HS}(\sigma) \, K_v \, F_\text{bond}.
\end{equation} 
Here, \( g_{HS}(\sigma) \) is the contact value of the pair distribution function for the hard sphere model, per Carnahan-Sterling:
\begin{equation} \label{eq:PairDistHSEta}
g_\text{HS} = \dfrac{1-\eta/2}{(1 - \eta)^3}.
\end{equation}
The quantity \( K_v \) is the available bonding volume on each particle, i.e., the effective volume around a given particle for the purpose of being captured by another particle to form a bond. The extent and shape of the binding potential enter the description through $K_v$.~\cite{JCG1988AL} In dimensionless terms, \(K_v^*=K_v/\sigma^3\). We adopt here a relatively small value \(K_v^*=0.5*10^{-4}\) consistent with having just one bonding site.~\cite{Talanquer2005phase} Lastly, \(F_\text{bond}\) is the Mayer $f$-function for the square-well binding potential:
\begin{equation}
F_\text{bond} = \exp(\epsilon_\text{bond}/k_B T) - 1.
\end{equation}

The quantity \( \lambda_\text{att}\) in Eq.~\eqref{eq:MassActionEquilibriumX}, by construction, incorporates effects of changes in dispersion interactions due to binding. Talanquer~\cite{Talanquer2005phase} derives the following approximation: \[ \lambda_\text{att} = \exp[-\alpha\sigma^3 (\Delta_{md}\rho_m + \delta_{md}\rho)/k_B T] \] thus yielding, in view of Eq.~(\ref{eq:Densities}):
\begin{equation} \label{eq:LambdaAttX}
\lambda_\text{att} = \exp[-\alpha\sigma^3\rho (\Delta_{md}X + \delta_{md})/k_B T].
\end{equation}
Alternatively, one may think of \( \lambda_\text{att}\) as correcting for the use of concentrations---as opposed to activities---in Eq.~\eqref{massact}, in the spirit of the venerable laws due to Henry and Raoult, c.f. Eq.~(A.4) of Ref.~\cite{HLKnowledge}

Combining equations (\ref{eq:MassActionEquilibriumX}) through (\ref{eq:LambdaAttX}) yields the following relation between the total density \(\rho\) and the monomer fraction \(X\):
\begin{align} \label{eq:RhoX} 1 - X &= \\  &  8 \pi \rho X^2 g_{HS}(\sigma)
 \, K_v \, F_\text{bond} \exp[-\alpha\sigma^3\rho (\Delta_{md}X + \delta_{md})/k_B T]. \nonumber
\end{align}

Equilibrium values of the free energy of the system are computed (numerically) by optimizing the free energy from Eq.~\eqref{eq:FreeEnergyFull} subject to the constraint due to Eq.~\eqref{eq:RhoX}. Note such equilibrium values may not be unique because Eq.~\eqref{eq:RhoX} allows for up to three distinct solutions.

\bibliography{lowT,Dissert}

\begin{thebibliography}{76}%
\makeatletter
\providecommand \@ifxundefined [1]{%
 \@ifx{#1\undefined}
}%
\providecommand \@ifnum [1]{%
 \ifnum #1\expandafter \@firstoftwo
 \else \expandafter \@secondoftwo
 \fi
}%
\providecommand \@ifx [1]{%
 \ifx #1\expandafter \@firstoftwo
 \else \expandafter \@secondoftwo
 \fi
}%
\providecommand \natexlab [1]{#1}%
\providecommand \enquote  [1]{``#1''}%
\providecommand \bibnamefont  [1]{#1}%
\providecommand \bibfnamefont [1]{#1}%
\providecommand \citenamefont [1]{#1}%
\providecommand \href@noop [0]{\@secondoftwo}%
\providecommand \href [0]{\begingroup \@sanitize@url \@href}%
\providecommand \@href[1]{\@@startlink{#1}\@@href}%
\providecommand \@@href[1]{\endgroup#1\@@endlink}%
\providecommand \@sanitize@url [0]{\catcode `\\12\catcode `\$12\catcode
  `\&12\catcode `\#12\catcode `\^12\catcode `\_12\catcode `\%12\relax}%
\providecommand \@@startlink[1]{}%
\providecommand \@@endlink[0]{}%
\providecommand \url  [0]{\begingroup\@sanitize@url \@url }%
\providecommand \@url [1]{\endgroup\@href {#1}{\urlprefix }}%
\providecommand \urlprefix  [0]{URL }%
\providecommand \Eprint [0]{\href }%
\providecommand \doibase [0]{http://dx.doi.org/}%
\providecommand \selectlanguage [0]{\@gobble}%
\providecommand \bibinfo  [0]{\@secondoftwo}%
\providecommand \bibfield  [0]{\@secondoftwo}%
\providecommand \translation [1]{[#1]}%
\providecommand \BibitemOpen [0]{}%
\providecommand \bibitemStop [0]{}%
\providecommand \bibitemNoStop [0]{.\EOS\space}%
\providecommand \EOS [0]{\spacefactor3000\relax}%
\providecommand \BibitemShut  [1]{\csname bibitem#1\endcsname}%
\let\auto@bib@innerbib\@empty
\bibitem [{\citenamefont {Gliko}\ \emph
  {et~al.}(2005{\natexlab{a}})\citenamefont {Gliko}, \citenamefont {Neumaier},
  \citenamefont {Pan}, \citenamefont {Haase}, \citenamefont {Fischer},
  \citenamefont {Bacher}, \citenamefont {Weinkauf},\ and\ \citenamefont
  {Vekilov}}]{GlikoJACS2005}%
  \BibitemOpen
  \bibfield  {author} {\bibinfo {author} {\bibfnamefont {O.}~\bibnamefont
  {Gliko}}, \bibinfo {author} {\bibfnamefont {N.}~\bibnamefont {Neumaier}},
  \bibinfo {author} {\bibfnamefont {W.}~\bibnamefont {Pan}}, \bibinfo {author}
  {\bibfnamefont {I.}~\bibnamefont {Haase}}, \bibinfo {author} {\bibfnamefont
  {M.}~\bibnamefont {Fischer}}, \bibinfo {author} {\bibfnamefont
  {A.}~\bibnamefont {Bacher}}, \bibinfo {author} {\bibfnamefont
  {S.}~\bibnamefont {Weinkauf}}, \ and\ \bibinfo {author} {\bibfnamefont
  {P.~G.}\ \bibnamefont {Vekilov}},\ }\bibfield  {title} {\enquote {\bibinfo
  {title} {A metastable prerequisite for the growth of lumazine synthase
  crystals},}\ }\href@noop {} {\bibfield  {journal} {\bibinfo  {journal}
  {JACS}\ }\textbf {\bibinfo {volume} {127}},\ \bibinfo {pages} {3433}
  (\bibinfo {year} {2005}{\natexlab{a}})}\BibitemShut {NoStop}%
\bibitem [{\citenamefont {Georgalis}\ \emph {et~al.}(1999)\citenamefont
  {Georgalis}, \citenamefont {Umbach}, \citenamefont {Saenger}, \citenamefont
  {Ihmels},\ and\ \citenamefont {Soumpasis}}]{Georgalis1999}%
  \BibitemOpen
  \bibfield  {author} {\bibinfo {author} {\bibfnamefont {Y.}~\bibnamefont
  {Georgalis}}, \bibinfo {author} {\bibfnamefont {P.}~\bibnamefont {Umbach}},
  \bibinfo {author} {\bibfnamefont {W.}~\bibnamefont {Saenger}}, \bibinfo
  {author} {\bibfnamefont {B.}~\bibnamefont {Ihmels}}, \ and\ \bibinfo {author}
  {\bibfnamefont {D.~M.}\ \bibnamefont {Soumpasis}},\ }\bibfield  {title}
  {\enquote {\bibinfo {title} {Ordering of fractal clusters in crystallizing
  lysozyme solutions},}\ }\href@noop {} {\bibfield  {journal} {\bibinfo
  {journal} {J. Amer. Chem. Soc.}\ }\textbf {\bibinfo {volume} {121}},\
  \bibinfo {pages} {1627--1635} (\bibinfo {year} {1999})}\BibitemShut {NoStop}%
\bibitem [{\citenamefont {Gliko}\ \emph {et~al.}(2007)\citenamefont {Gliko},
  \citenamefont {Pan}, \citenamefont {Katsonis}, \citenamefont {Neumaier},
  \citenamefont {Galkin}, \citenamefont {Weinkauf},\ and\ \citenamefont
  {Vekilov}}]{doi:10.1021/jp068827o}%
  \BibitemOpen
  \bibfield  {author} {\bibinfo {author} {\bibfnamefont {O.}~\bibnamefont
  {Gliko}}, \bibinfo {author} {\bibfnamefont {W.}~\bibnamefont {Pan}}, \bibinfo
  {author} {\bibfnamefont {P.}~\bibnamefont {Katsonis}}, \bibinfo {author}
  {\bibfnamefont {N.}~\bibnamefont {Neumaier}}, \bibinfo {author}
  {\bibfnamefont {O.}~\bibnamefont {Galkin}}, \bibinfo {author} {\bibfnamefont
  {S.}~\bibnamefont {Weinkauf}}, \ and\ \bibinfo {author} {\bibfnamefont
  {P.~G.}\ \bibnamefont {Vekilov}},\ }\bibfield  {title} {\enquote {\bibinfo
  {title} {Metastable liquid clusters in super- and undersaturated protein
  solutions},}\ }\href@noop {} {\bibfield  {journal} {\bibinfo  {journal} {J.
  Phys. Chem. B}\ }\textbf {\bibinfo {volume} {111}},\ \bibinfo {pages}
  {3106--3114} (\bibinfo {year} {2007})}\BibitemShut {NoStop}%
\bibitem [{\citenamefont {Pan}, \citenamefont {Vekilov},\ and\ \citenamefont
  {Lubchenko}(2010)}]{PVL}%
  \BibitemOpen
  \bibfield  {author} {\bibinfo {author} {\bibfnamefont {W.}~\bibnamefont
  {Pan}}, \bibinfo {author} {\bibfnamefont {P.~G.}\ \bibnamefont {Vekilov}}, \
  and\ \bibinfo {author} {\bibfnamefont {V.}~\bibnamefont {Lubchenko}},\
  }\bibfield  {title} {\enquote {\bibinfo {title} {Origin of anomalous
  mesoscopic phases in protein solutions},}\ }\href@noop {} {\bibfield
  {journal} {\bibinfo  {journal} {J. Phys. Chem. B}\ }\textbf {\bibinfo
  {volume} {114}},\ \bibinfo {pages} {7620--7630} (\bibinfo {year}
  {2010})}\BibitemShut {NoStop}%
\bibitem [{\citenamefont {Li}\ \emph {et~al.}(2012)\citenamefont {Li},
  \citenamefont {Lubchenko}, \citenamefont {Vorontsova}, \citenamefont
  {Filobelo},\ and\ \citenamefont {Vekilov}}]{LLVFPostwaldRipening}%
  \BibitemOpen
  \bibfield  {author} {\bibinfo {author} {\bibfnamefont {Y.}~\bibnamefont
  {Li}}, \bibinfo {author} {\bibfnamefont {V.}~\bibnamefont {Lubchenko}},
  \bibinfo {author} {\bibfnamefont {M.~A.}\ \bibnamefont {Vorontsova}},
  \bibinfo {author} {\bibfnamefont {L.}~\bibnamefont {Filobelo}}, \ and\
  \bibinfo {author} {\bibfnamefont {P.~G.}\ \bibnamefont {Vekilov}},\
  }\bibfield  {title} {\enquote {\bibinfo {title} {Ostwald-like ripening of the
  anomalous mesoscopic clusters in protein solutions},}\ }\href@noop {}
  {\bibfield  {journal} {\bibinfo  {journal} {J. Phys. Chem. B}\ }\textbf
  {\bibinfo {volume} {116}},\ \bibinfo {pages} {10657--10664} (\bibinfo {year}
  {2012})}\BibitemShut {NoStop}%
\bibitem [{\citenamefont {Yearley}\ \emph {et~al.}(2014)\citenamefont
  {Yearley}, \citenamefont {Godfrin}, \citenamefont {Perevozchikova},
  \citenamefont {Zhang}, \citenamefont {Falus}, \citenamefont {Porcar},
  \citenamefont {Nagao}, \citenamefont {Curtis}, \citenamefont {Gawande},
  \citenamefont {Taing}, \citenamefont {Zarraga}, \citenamefont {Wagner},\ and\
  \citenamefont {Liu}}]{YEARLEY20141763}%
  \BibitemOpen
  \bibfield  {author} {\bibinfo {author} {\bibfnamefont {E.~J.}\ \bibnamefont
  {Yearley}}, \bibinfo {author} {\bibfnamefont {P.~D.}\ \bibnamefont
  {Godfrin}}, \bibinfo {author} {\bibfnamefont {T.}~\bibnamefont
  {Perevozchikova}}, \bibinfo {author} {\bibfnamefont {H.}~\bibnamefont
  {Zhang}}, \bibinfo {author} {\bibfnamefont {P.}~\bibnamefont {Falus}},
  \bibinfo {author} {\bibfnamefont {L.}~\bibnamefont {Porcar}}, \bibinfo
  {author} {\bibfnamefont {M.}~\bibnamefont {Nagao}}, \bibinfo {author}
  {\bibfnamefont {J.~E.}\ \bibnamefont {Curtis}}, \bibinfo {author}
  {\bibfnamefont {P.}~\bibnamefont {Gawande}}, \bibinfo {author} {\bibfnamefont
  {R.}~\bibnamefont {Taing}}, \bibinfo {author} {\bibfnamefont {I.~E.}\
  \bibnamefont {Zarraga}}, \bibinfo {author} {\bibfnamefont {N.~J.}\
  \bibnamefont {Wagner}}, \ and\ \bibinfo {author} {\bibfnamefont
  {Y.}~\bibnamefont {Liu}},\ }\bibfield  {title} {\enquote {\bibinfo {title}
  {Observation of small cluster formation in concentrated monoclonal antibody
  solutions and its implications to solution viscosity},}\ }\href@noop {}
  {\bibfield  {journal} {\bibinfo  {journal} {Biophys. J.}\ }\textbf {\bibinfo
  {volume} {106}},\ \bibinfo {pages} {1763--1770} (\bibinfo {year}
  {2014})}\BibitemShut {NoStop}%
\bibitem [{\citenamefont {Sleutel}\ and\ \citenamefont
  {Van~Driessche}(2014)}]{SleutelE546}%
  \BibitemOpen
  \bibfield  {author} {\bibinfo {author} {\bibfnamefont {M.}~\bibnamefont
  {Sleutel}}\ and\ \bibinfo {author} {\bibfnamefont {A.~E.~S.}\ \bibnamefont
  {Van~Driessche}},\ }\bibfield  {title} {\enquote {\bibinfo {title} {Role of
  clusters in nonclassical nucleation and growth of protein crystals},}\
  }\href@noop {} {\bibfield  {journal} {\bibinfo  {journal} {Proc. Natl. Acad.
  Sci. U.~S.~A.}\ }\textbf {\bibinfo {volume} {111}},\ \bibinfo {pages}
  {E546--E553} (\bibinfo {year} {2014})}\BibitemShut {NoStop}%
\bibitem [{\citenamefont {Schubert}\ \emph {et~al.}(2017)\citenamefont
  {Schubert}, \citenamefont {Meyer}, \citenamefont {Baitan}, \citenamefont
  {Dierks}, \citenamefont {Perbandt},\ and\ \citenamefont
  {Betzel}}]{doi:10.1021/acs.cgd.6b01826}%
  \BibitemOpen
  \bibfield  {author} {\bibinfo {author} {\bibfnamefont {R.}~\bibnamefont
  {Schubert}}, \bibinfo {author} {\bibfnamefont {A.}~\bibnamefont {Meyer}},
  \bibinfo {author} {\bibfnamefont {D.}~\bibnamefont {Baitan}}, \bibinfo
  {author} {\bibfnamefont {K.}~\bibnamefont {Dierks}}, \bibinfo {author}
  {\bibfnamefont {M.}~\bibnamefont {Perbandt}}, \ and\ \bibinfo {author}
  {\bibfnamefont {C.}~\bibnamefont {Betzel}},\ }\bibfield  {title} {\enquote
  {\bibinfo {title} {Real-time observation of protein dense liquid cluster
  evolution during nucleation in protein crystallization},}\ }\href@noop {}
  {\bibfield  {journal} {\bibinfo  {journal} {Crystal Growth \& Design}\
  }\textbf {\bibinfo {volume} {17}},\ \bibinfo {pages} {954--958} (\bibinfo
  {year} {2017})}\BibitemShut {NoStop}%
\bibitem [{\citenamefont {Gliko}\ \emph
  {et~al.}(2005{\natexlab{b}})\citenamefont {Gliko}, \citenamefont {Neumaier},
  \citenamefont {Pan}, \citenamefont {Haase}, \citenamefont {Fischer},
  \citenamefont {Bacher}, \citenamefont {Weinkauf},\ and\ \citenamefont
  {Vekilov}}]{Gliko2005}%
  \BibitemOpen
  \bibfield  {author} {\bibinfo {author} {\bibfnamefont {O.}~\bibnamefont
  {Gliko}}, \bibinfo {author} {\bibfnamefont {N.}~\bibnamefont {Neumaier}},
  \bibinfo {author} {\bibfnamefont {W.}~\bibnamefont {Pan}}, \bibinfo {author}
  {\bibfnamefont {I.}~\bibnamefont {Haase}}, \bibinfo {author} {\bibfnamefont
  {M.}~\bibnamefont {Fischer}}, \bibinfo {author} {\bibfnamefont
  {A.}~\bibnamefont {Bacher}}, \bibinfo {author} {\bibfnamefont
  {S.}~\bibnamefont {Weinkauf}}, \ and\ \bibinfo {author} {\bibfnamefont
  {P.~G.}\ \bibnamefont {Vekilov}},\ }\bibfield  {title} {\enquote {\bibinfo
  {title} {Dense liquid droplets as a step source for the crystallization of
  lumazine synthase},}\ }\href@noop {} {\bibfield  {journal} {\bibinfo
  {journal} {Journal of Cryst. Growth}\ }\textbf {\bibinfo {volume} {275}},\
  \bibinfo {pages} {e1409} (\bibinfo {year} {2005}{\natexlab{b}})}\BibitemShut
  {NoStop}%
\bibitem [{\citenamefont {Yamazaki}\ \emph {et~al.}(2017)\citenamefont
  {Yamazaki}, \citenamefont {Kimura}, \citenamefont {Vekilov}, \citenamefont
  {Furukawa}, \citenamefont {Shirai}, \citenamefont {Matsumoto}, \citenamefont
  {Van~Driessche},\ and\ \citenamefont {Tsukamoto}}]{Yamazaki2154}%
  \BibitemOpen
  \bibfield  {author} {\bibinfo {author} {\bibfnamefont {T.}~\bibnamefont
  {Yamazaki}}, \bibinfo {author} {\bibfnamefont {Y.}~\bibnamefont {Kimura}},
  \bibinfo {author} {\bibfnamefont {P.~G.}\ \bibnamefont {Vekilov}}, \bibinfo
  {author} {\bibfnamefont {E.}~\bibnamefont {Furukawa}}, \bibinfo {author}
  {\bibfnamefont {M.}~\bibnamefont {Shirai}}, \bibinfo {author} {\bibfnamefont
  {H.}~\bibnamefont {Matsumoto}}, \bibinfo {author} {\bibfnamefont {A.~E.~S.}\
  \bibnamefont {Van~Driessche}}, \ and\ \bibinfo {author} {\bibfnamefont
  {K.}~\bibnamefont {Tsukamoto}},\ }\bibfield  {title} {\enquote {\bibinfo
  {title} {Two types of amorphous protein particles facilitate crystal
  nucleation},}\ }\href@noop {} {\bibfield  {journal} {\bibinfo  {journal}
  {Proc. Natl. Acad. Sci. U.~S.~A.}\ }\textbf {\bibinfo {volume} {114}},\
  \bibinfo {pages} {2154--2159} (\bibinfo {year} {2017})}\BibitemShut {NoStop}%
\bibitem [{\citenamefont {Safari}\ \emph {et~al.}(2019)\citenamefont {Safari},
  \citenamefont {Wang}, \citenamefont {Tailor}, \citenamefont {Kolomeisky},
  \citenamefont {Conrad},\ and\ \citenamefont {Vekilov}}]{p53clusters}%
  \BibitemOpen
  \bibfield  {author} {\bibinfo {author} {\bibfnamefont {M.~S.}\ \bibnamefont
  {Safari}}, \bibinfo {author} {\bibfnamefont {Z.}~\bibnamefont {Wang}},
  \bibinfo {author} {\bibfnamefont {K.}~\bibnamefont {Tailor}}, \bibinfo
  {author} {\bibfnamefont {A.~B.}\ \bibnamefont {Kolomeisky}}, \bibinfo
  {author} {\bibfnamefont {J.~C.}\ \bibnamefont {Conrad}}, \ and\ \bibinfo
  {author} {\bibfnamefont {P.~G.}\ \bibnamefont {Vekilov}},\ }\bibfield
  {title} {\enquote {\bibinfo {title} {Anomalous dense liquid condensates host
  the nucleation of tumor suppressor p53 fibrils},}\ }\href@noop {} {\bibfield
  {journal} {\bibinfo  {journal} {iScience}\ }\textbf {\bibinfo {volume}
  {12}},\ \bibinfo {pages} {342--355} (\bibinfo {year} {2019})}\BibitemShut
  {NoStop}%
\bibitem [{\citenamefont {Kar}\ \emph {et~al.}(2022)\citenamefont {Kar},
  \citenamefont {Dar}, \citenamefont {Welsh}, \citenamefont {Vogel},
  \citenamefont {Kühnemuth}, \citenamefont {Majumdar}, \citenamefont
  {Krainer}, \citenamefont {Franzmann}, \citenamefont {Alberti}, \citenamefont
  {Seidel}, \citenamefont {Knowles}, \citenamefont {Hyman},\ and\ \citenamefont
  {Pappu}}]{doi:10.1073/pnas.2202222119}%
  \BibitemOpen
  \bibfield  {author} {\bibinfo {author} {\bibfnamefont {M.}~\bibnamefont
  {Kar}}, \bibinfo {author} {\bibfnamefont {F.}~\bibnamefont {Dar}}, \bibinfo
  {author} {\bibfnamefont {T.~J.}\ \bibnamefont {Welsh}}, \bibinfo {author}
  {\bibfnamefont {L.~T.}\ \bibnamefont {Vogel}}, \bibinfo {author}
  {\bibfnamefont {R.}~\bibnamefont {Kühnemuth}}, \bibinfo {author}
  {\bibfnamefont {A.}~\bibnamefont {Majumdar}}, \bibinfo {author}
  {\bibfnamefont {G.}~\bibnamefont {Krainer}}, \bibinfo {author} {\bibfnamefont
  {T.~M.}\ \bibnamefont {Franzmann}}, \bibinfo {author} {\bibfnamefont
  {S.}~\bibnamefont {Alberti}}, \bibinfo {author} {\bibfnamefont {C.~A.~M.}\
  \bibnamefont {Seidel}}, \bibinfo {author} {\bibfnamefont {T.~P.~J.}\
  \bibnamefont {Knowles}}, \bibinfo {author} {\bibfnamefont {A.~A.}\
  \bibnamefont {Hyman}}, \ and\ \bibinfo {author} {\bibfnamefont {R.~V.}\
  \bibnamefont {Pappu}},\ }\bibfield  {title} {\enquote {\bibinfo {title}
  {Phase-separating rna-binding proteins form heterogeneous distributions of
  clusters in subsaturated solutions},}\ }\href@noop {} {\bibfield  {journal}
  {\bibinfo  {journal} {Proceedings of the National Academy of Sciences}\
  }\textbf {\bibinfo {volume} {119}},\ \bibinfo {pages} {e2202222119} (\bibinfo
  {year} {2022})}\BibitemShut {NoStop}%
\bibitem [{\citenamefont {Li}, \citenamefont {Lubchenko},\ and\ \citenamefont
  {Vekilov}(2011)}]{10.1063/1.3592581}%
  \BibitemOpen
  \bibfield  {author} {\bibinfo {author} {\bibfnamefont {Y.}~\bibnamefont
  {Li}}, \bibinfo {author} {\bibfnamefont {V.}~\bibnamefont {Lubchenko}}, \
  and\ \bibinfo {author} {\bibfnamefont {P.~G.}\ \bibnamefont {Vekilov}},\
  }\bibfield  {title} {\enquote {\bibinfo {title} {The use of dynamic light
  scattering and brownian microscopy to characterize protein aggregation},}\
  }\href@noop {} {\bibfield  {journal} {\bibinfo  {journal} {Rev. Sci. Instr.}\
  }\textbf {\bibinfo {volume} {82}},\ \bibinfo {pages} {053106} (\bibinfo
  {year} {2011})}\BibitemShut {NoStop}%
\bibitem [{\citenamefont {Pan}\ \emph {et~al.}(2007)\citenamefont {Pan},
  \citenamefont {Galkin}, \citenamefont {Filobelo}, \citenamefont {Nagel},\
  and\ \citenamefont {Vekilov}}]{PAN2007267}%
  \BibitemOpen
  \bibfield  {author} {\bibinfo {author} {\bibfnamefont {W.}~\bibnamefont
  {Pan}}, \bibinfo {author} {\bibfnamefont {O.}~\bibnamefont {Galkin}},
  \bibinfo {author} {\bibfnamefont {L.}~\bibnamefont {Filobelo}}, \bibinfo
  {author} {\bibfnamefont {R.~L.}\ \bibnamefont {Nagel}}, \ and\ \bibinfo
  {author} {\bibfnamefont {P.~G.}\ \bibnamefont {Vekilov}},\ }\bibfield
  {title} {\enquote {\bibinfo {title} {Metastable mesoscopic clusters in
  solutions of sickle-cell hemoglobin},}\ }\href@noop {} {\bibfield  {journal}
  {\bibinfo  {journal} {Biophysical Journal}\ }\textbf {\bibinfo {volume}
  {92}},\ \bibinfo {pages} {267--277} (\bibinfo {year} {2007})}\BibitemShut
  {NoStop}%
\bibitem [{\citenamefont {Muschol}\ and\ \citenamefont
  {Rosenberger}(1997)}]{Muschol1997}%
  \BibitemOpen
  \bibfield  {author} {\bibinfo {author} {\bibfnamefont {M.}~\bibnamefont
  {Muschol}}\ and\ \bibinfo {author} {\bibfnamefont {F.}~\bibnamefont
  {Rosenberger}},\ }\bibfield  {title} {\enquote {\bibinfo {title}
  {Liquid-liquid phase separation in supersaturated lysozyme solutions and
  associated precipitate formation/crystallization},}\ }\href@noop {}
  {\bibfield  {journal} {\bibinfo  {journal} {J. Chem. Phys.}\ }\textbf
  {\bibinfo {volume} {107}},\ \bibinfo {pages} {1953--1962} (\bibinfo {year}
  {1997})}\BibitemShut {NoStop}%
\bibitem [{\citenamefont {{ten Wolde}}\ and\ \citenamefont
  {Frenkel}(1997)}]{WoldeFrenkel}%
  \BibitemOpen
  \bibfield  {author} {\bibinfo {author} {\bibfnamefont {P.~R.}\ \bibnamefont
  {{ten Wolde}}}\ and\ \bibinfo {author} {\bibfnamefont {D.}~\bibnamefont
  {Frenkel}},\ }\bibfield  {title} {\enquote {\bibinfo {title} {Enhancement of
  protein crystal nucleation by critical density fluctuations},}\ }\href@noop
  {} {\bibfield  {journal} {\bibinfo  {journal} {Science}\ }\textbf {\bibinfo
  {volume} {277}},\ \bibinfo {pages} {1975--1978} (\bibinfo {year}
  {1997})}\BibitemShut {NoStop}%
\bibitem [{\citenamefont {Harrington}\ \emph {et~al.}(1997)\citenamefont
  {Harrington}, \citenamefont {Zhang}, \citenamefont {Poole}, \citenamefont
  {Sciortino},\ and\ \citenamefont {Stanley}}]{PhysRevLett.78.2409}%
  \BibitemOpen
  \bibfield  {author} {\bibinfo {author} {\bibfnamefont {S.}~\bibnamefont
  {Harrington}}, \bibinfo {author} {\bibfnamefont {R.}~\bibnamefont {Zhang}},
  \bibinfo {author} {\bibfnamefont {P.~H.}\ \bibnamefont {Poole}}, \bibinfo
  {author} {\bibfnamefont {F.}~\bibnamefont {Sciortino}}, \ and\ \bibinfo
  {author} {\bibfnamefont {H.~E.}\ \bibnamefont {Stanley}},\ }\bibfield
  {title} {\enquote {\bibinfo {title} {Liquid-liquid phase transition: Evidence
  from simulations},}\ }\href@noop {} {\bibfield  {journal} {\bibinfo
  {journal} {Phys. Rev. Lett.}\ }\textbf {\bibinfo {volume} {78}},\ \bibinfo
  {pages} {2409--2412} (\bibinfo {year} {1997})}\BibitemShut {NoStop}%
\bibitem [{\citenamefont {Yang}\ \emph {et~al.}(2021)\citenamefont {Yang},
  \citenamefont {Saeedi}, \citenamefont {Davtyan}, \citenamefont {Fathi},
  \citenamefont {Sherman}, \citenamefont {Safari}, \citenamefont {Klindziuk},
  \citenamefont {Barton}, \citenamefont {Varadarajan}, \citenamefont
  {Kolomeisky},\ and\ \citenamefont {Vekilov}}]{Yange2015618118}%
  \BibitemOpen
  \bibfield  {author} {\bibinfo {author} {\bibfnamefont {D.~S.}\ \bibnamefont
  {Yang}}, \bibinfo {author} {\bibfnamefont {A.}~\bibnamefont {Saeedi}},
  \bibinfo {author} {\bibfnamefont {A.}~\bibnamefont {Davtyan}}, \bibinfo
  {author} {\bibfnamefont {M.}~\bibnamefont {Fathi}}, \bibinfo {author}
  {\bibfnamefont {M.~B.}\ \bibnamefont {Sherman}}, \bibinfo {author}
  {\bibfnamefont {M.~S.}\ \bibnamefont {Safari}}, \bibinfo {author}
  {\bibfnamefont {A.}~\bibnamefont {Klindziuk}}, \bibinfo {author}
  {\bibfnamefont {M.~C.}\ \bibnamefont {Barton}}, \bibinfo {author}
  {\bibfnamefont {N.}~\bibnamefont {Varadarajan}}, \bibinfo {author}
  {\bibfnamefont {A.~B.}\ \bibnamefont {Kolomeisky}}, \ and\ \bibinfo {author}
  {\bibfnamefont {P.~G.}\ \bibnamefont {Vekilov}},\ }\bibfield  {title}
  {\enquote {\bibinfo {title} {Mesoscopic protein-rich clusters host the
  nucleation of mutant p53 amyloid fibrils},}\ }\href@noop {} {\bibfield
  {journal} {\bibinfo  {journal} {Proc. Natl. Acad. Sci. U.~S.~A.}\ }\textbf
  {\bibinfo {volume} {118}} (\bibinfo {year} {2021})}\BibitemShut {NoStop}%
\bibitem [{\citenamefont {Aich}, \citenamefont {Pan},\ and\ \citenamefont
  {Vekilov}(2015)}]{https://doi.org/10.1002/aic.14800}%
  \BibitemOpen
  \bibfield  {author} {\bibinfo {author} {\bibfnamefont {A.}~\bibnamefont
  {Aich}}, \bibinfo {author} {\bibfnamefont {W.}~\bibnamefont {Pan}}, \ and\
  \bibinfo {author} {\bibfnamefont {P.~G.}\ \bibnamefont {Vekilov}},\
  }\bibfield  {title} {\enquote {\bibinfo {title} {Thermodynamic mechanism of
  free heme action on sickle cell hemoglobin polymerization},}\ }\href@noop {}
  {\bibfield  {journal} {\bibinfo  {journal} {AIChE Journal}\ }\textbf
  {\bibinfo {volume} {61}},\ \bibinfo {pages} {2861--2870} (\bibinfo {year}
  {2015})}\BibitemShut {NoStop}%
\bibitem [{\citenamefont {Uzunova}\ \emph {et~al.}(2010)\citenamefont
  {Uzunova}, \citenamefont {Pan}, \citenamefont {Galkin},\ and\ \citenamefont
  {Vekilov}}]{UZUNOVA20101976}%
  \BibitemOpen
  \bibfield  {author} {\bibinfo {author} {\bibfnamefont {V.~V.}\ \bibnamefont
  {Uzunova}}, \bibinfo {author} {\bibfnamefont {W.}~\bibnamefont {Pan}},
  \bibinfo {author} {\bibfnamefont {O.}~\bibnamefont {Galkin}}, \ and\ \bibinfo
  {author} {\bibfnamefont {P.~G.}\ \bibnamefont {Vekilov}},\ }\bibfield
  {title} {\enquote {\bibinfo {title} {Free heme and the polymerization of
  sickle cell hemoglobin},}\ }\href@noop {} {\bibfield  {journal} {\bibinfo
  {journal} {Biophys. J.}\ }\textbf {\bibinfo {volume} {99}},\ \bibinfo {pages}
  {1976 -- 1985} (\bibinfo {year} {2010})}\BibitemShut {NoStop}%
\bibitem [{\citenamefont {Warzecha}\ \emph {et~al.}(2017)\citenamefont
  {Warzecha}, \citenamefont {Safari}, \citenamefont {Florence},\ and\
  \citenamefont {Vekilov}}]{doi:10.1021/acs.cgd.7b01299}%
  \BibitemOpen
  \bibfield  {author} {\bibinfo {author} {\bibfnamefont {M.}~\bibnamefont
  {Warzecha}}, \bibinfo {author} {\bibfnamefont {M.~S.}\ \bibnamefont
  {Safari}}, \bibinfo {author} {\bibfnamefont {A.~J.}\ \bibnamefont
  {Florence}}, \ and\ \bibinfo {author} {\bibfnamefont {P.~G.}\ \bibnamefont
  {Vekilov}},\ }\bibfield  {title} {\enquote {\bibinfo {title} {Mesoscopic
  solute-rich clusters in olanzapine solutions},}\ }\href@noop {} {\bibfield
  {journal} {\bibinfo  {journal} {Crystal Growth \& Design}\ }\textbf {\bibinfo
  {volume} {17}},\ \bibinfo {pages} {6668--6676} (\bibinfo {year}
  {2017})}\BibitemShut {NoStop}%
\bibitem [{\citenamefont {Chakraborty}\ \emph {et~al.}(2025)\citenamefont
  {Chakraborty}, \citenamefont {Ma}, \citenamefont {Vekilov},\ and\
  \citenamefont {Rimer}}]{doi:10.1021/acs.chemmater.5c00679}%
  \BibitemOpen
  \bibfield  {author} {\bibinfo {author} {\bibfnamefont {D.}~\bibnamefont
  {Chakraborty}}, \bibinfo {author} {\bibfnamefont {W.}~\bibnamefont {Ma}},
  \bibinfo {author} {\bibfnamefont {P.~G.}\ \bibnamefont {Vekilov}}, \ and\
  \bibinfo {author} {\bibfnamefont {J.~D.}\ \bibnamefont {Rimer}},\ }\bibfield
  {title} {\enquote {\bibinfo {title} {Role of mesoscopic solute-rich clusters
  in cholesterol crystallization},}\ }\href@noop {} {\bibfield  {journal}
  {\bibinfo  {journal} {Chemistry of Materials}\ }\textbf {\bibinfo {volume}
  {37}},\ \bibinfo {pages} {4158--4168} (\bibinfo {year} {2025})}\BibitemShut
  {NoStop}%
\bibitem [{\citenamefont {Yerragunta}\ \emph {et~al.}(2025)\citenamefont
  {Yerragunta}, \citenamefont {Veliz}, \citenamefont {Warzecha}, \citenamefont
  {Hadjiev}, \citenamefont {Florence}, \citenamefont {Zerze}, \citenamefont
  {Rimer},\ and\ \citenamefont {Vekilov}}]{doi:10.1021/acs.cgd.5c00370}%
  \BibitemOpen
  \bibfield  {author} {\bibinfo {author} {\bibfnamefont {M.}~\bibnamefont
  {Yerragunta}}, \bibinfo {author} {\bibfnamefont {A.~C.}\ \bibnamefont
  {Veliz}}, \bibinfo {author} {\bibfnamefont {M.}~\bibnamefont {Warzecha}},
  \bibinfo {author} {\bibfnamefont {V.~G.}\ \bibnamefont {Hadjiev}}, \bibinfo
  {author} {\bibfnamefont {A.~J.}\ \bibnamefont {Florence}}, \bibinfo {author}
  {\bibfnamefont {G.~H.}\ \bibnamefont {Zerze}}, \bibinfo {author}
  {\bibfnamefont {J.~D.}\ \bibnamefont {Rimer}}, \ and\ \bibinfo {author}
  {\bibfnamefont {P.~G.}\ \bibnamefont {Vekilov}},\ }\bibfield  {title}
  {\enquote {\bibinfo {title} {Mesoscopic solute-rich clusters in organic
  solutions},}\ }\href@noop {} {\bibfield  {journal} {\bibinfo  {journal}
  {Crystal Growth \& Design}\ }\textbf {\bibinfo {volume} {25}},\ \bibinfo
  {pages} {3958--3967} (\bibinfo {year} {2025})}\BibitemShut {NoStop}%
\bibitem [{\citenamefont {Chan}\ and\ \citenamefont
  {Lubchenko}(2019)}]{CL_NatureComm}%
  \BibitemOpen
  \bibfield  {author} {\bibinfo {author} {\bibfnamefont {H.~Y.}\ \bibnamefont
  {Chan}}\ and\ \bibinfo {author} {\bibfnamefont {V.}~\bibnamefont
  {Lubchenko}},\ }\bibfield  {title} {\enquote {\bibinfo {title} {{A mechanism
  for reversible mesoscopic aggregation in liquid solutions}},}\ }\href@noop {}
  {\bibfield  {journal} {\bibinfo  {journal} {Nature Comm.}\ }\textbf {\bibinfo
  {volume} {10}},\ \bibinfo {pages} {2381} (\bibinfo {year}
  {2019})}\BibitemShut {NoStop}%
\bibitem [{\citenamefont {Shin}\ and\ \citenamefont
  {Brangwynne}(2017)}]{Shineaaf4382}%
  \BibitemOpen
  \bibfield  {author} {\bibinfo {author} {\bibfnamefont {Y.}~\bibnamefont
  {Shin}}\ and\ \bibinfo {author} {\bibfnamefont {C.~P.}\ \bibnamefont
  {Brangwynne}},\ }\bibfield  {title} {\enquote {\bibinfo {title} {Liquid phase
  condensation in cell physiology and disease},}\ }\href@noop {} {\bibfield
  {journal} {\bibinfo  {journal} {Science}\ }\textbf {\bibinfo {volume} {357}}
  (\bibinfo {year} {2017})}\BibitemShut {NoStop}%
\bibitem [{\citenamefont {Dai}\ \emph {et~al.}(2024)\citenamefont {Dai},
  \citenamefont {Zhou}, \citenamefont {Yu}, \citenamefont {Ma}, \citenamefont
  {Kim}, \citenamefont {Rivera}, \citenamefont {Mohammed}, \citenamefont
  {Lantelme}, \citenamefont {Hsu-Kim}, \citenamefont {Chilkoti},\ and\
  \citenamefont {You}}]{Dai2024}%
  \BibitemOpen
  \bibfield  {author} {\bibinfo {author} {\bibfnamefont {Y.}~\bibnamefont
  {Dai}}, \bibinfo {author} {\bibfnamefont {Z.}~\bibnamefont {Zhou}}, \bibinfo
  {author} {\bibfnamefont {W.}~\bibnamefont {Yu}}, \bibinfo {author}
  {\bibfnamefont {Y.}~\bibnamefont {Ma}}, \bibinfo {author} {\bibfnamefont
  {K.}~\bibnamefont {Kim}}, \bibinfo {author} {\bibfnamefont {N.}~\bibnamefont
  {Rivera}}, \bibinfo {author} {\bibfnamefont {J.}~\bibnamefont {Mohammed}},
  \bibinfo {author} {\bibfnamefont {E.}~\bibnamefont {Lantelme}}, \bibinfo
  {author} {\bibfnamefont {H.}~\bibnamefont {Hsu-Kim}}, \bibinfo {author}
  {\bibfnamefont {A.}~\bibnamefont {Chilkoti}}, \ and\ \bibinfo {author}
  {\bibfnamefont {L.}~\bibnamefont {You}},\ }\bibfield  {title} {\enquote
  {\bibinfo {title} {Biomolecular condensates regulate cellular electrochemical
  equilibria},}\ }\href@noop {} {\bibfield  {journal} {\bibinfo  {journal}
  {Cell}\ }\textbf {\bibinfo {volume} {187}},\ \bibinfo {pages}
  {5951--5966.e18} (\bibinfo {year} {2024})}\BibitemShut {NoStop}%
\bibitem [{\citenamefont {Smokers}\ \emph {et~al.}(2024)\citenamefont
  {Smokers}, \citenamefont {Visser}, \citenamefont {Slootbeek}, \citenamefont
  {Huck},\ and\ \citenamefont {Spruijt}}]{doi:10.1021/acs.accounts.4c00114}%
  \BibitemOpen
  \bibfield  {author} {\bibinfo {author} {\bibfnamefont {I.~B.~A.}\
  \bibnamefont {Smokers}}, \bibinfo {author} {\bibfnamefont {B.~S.}\
  \bibnamefont {Visser}}, \bibinfo {author} {\bibfnamefont {A.~D.}\
  \bibnamefont {Slootbeek}}, \bibinfo {author} {\bibfnamefont {W.~T.~S.}\
  \bibnamefont {Huck}}, \ and\ \bibinfo {author} {\bibfnamefont
  {E.}~\bibnamefont {Spruijt}},\ }\bibfield  {title} {\enquote {\bibinfo
  {title} {How droplets can accelerate reactions-coacervate protocells as
  catalytic microcompartments},}\ }\href@noop {} {\bibfield  {journal}
  {\bibinfo  {journal} {Acc. Chem. Res.}\ }\textbf {\bibinfo {volume} {57}},\
  \bibinfo {pages} {1885--1895} (\bibinfo {year} {2024})}\BibitemShut {NoStop}%
\bibitem [{\citenamefont {Zhou}\ \emph {et~al.}(2024)\citenamefont {Zhou},
  \citenamefont {Kota}, \citenamefont {Qin},\ and\ \citenamefont
  {Prasad}}]{doi:10.1021/acs.chemrev.4c00138}%
  \BibitemOpen
  \bibfield  {author} {\bibinfo {author} {\bibfnamefont {H.-X.}\ \bibnamefont
  {Zhou}}, \bibinfo {author} {\bibfnamefont {D.}~\bibnamefont {Kota}}, \bibinfo
  {author} {\bibfnamefont {S.}~\bibnamefont {Qin}}, \ and\ \bibinfo {author}
  {\bibfnamefont {R.}~\bibnamefont {Prasad}},\ }\bibfield  {title} {\enquote
  {\bibinfo {title} {Fundamental aspects of phase-separated biomolecular
  condensates},}\ }\href@noop {} {\bibfield  {journal} {\bibinfo  {journal}
  {Chemical Reviews}\ }\textbf {\bibinfo {volume} {124}},\ \bibinfo {pages}
  {8550--8595} (\bibinfo {year} {2024})}\BibitemShut {NoStop}%
\bibitem [{\citenamefont {Posey}\ \emph {et~al.}(2024)\citenamefont {Posey},
  \citenamefont {Bremer}, \citenamefont {Erkamp}, \citenamefont {Pant},
  \citenamefont {Knowles}, \citenamefont {Dai}, \citenamefont {Mittag},\ and\
  \citenamefont {Pappu}}]{doi:10.1021/jacs.4c08946}%
  \BibitemOpen
  \bibfield  {author} {\bibinfo {author} {\bibfnamefont {A.~E.}\ \bibnamefont
  {Posey}}, \bibinfo {author} {\bibfnamefont {A.}~\bibnamefont {Bremer}},
  \bibinfo {author} {\bibfnamefont {N.~A.}\ \bibnamefont {Erkamp}}, \bibinfo
  {author} {\bibfnamefont {A.}~\bibnamefont {Pant}}, \bibinfo {author}
  {\bibfnamefont {T.~P.~J.}\ \bibnamefont {Knowles}}, \bibinfo {author}
  {\bibfnamefont {Y.}~\bibnamefont {Dai}}, \bibinfo {author} {\bibfnamefont
  {T.}~\bibnamefont {Mittag}}, \ and\ \bibinfo {author} {\bibfnamefont {R.~V.}\
  \bibnamefont {Pappu}},\ }\bibfield  {title} {\enquote {\bibinfo {title}
  {Biomolecular condensates are characterized by interphase electric
  potentials},}\ }\href@noop {} {\bibfield  {journal} {\bibinfo  {journal}
  {JACS}\ }\textbf {\bibinfo {volume} {146}},\ \bibinfo {pages} {28268--28281}
  (\bibinfo {year} {2024})}\BibitemShut {NoStop}%
\bibitem [{\citenamefont {Bray}(1994)}]{Bray}%
  \BibitemOpen
  \bibfield  {author} {\bibinfo {author} {\bibfnamefont {A.~J.}\ \bibnamefont
  {Bray}},\ }\bibfield  {title} {\enquote {\bibinfo {title} {Theory of
  phase-ordering kinetics},}\ }\href@noop {} {\bibfield  {journal} {\bibinfo
  {journal} {Adv. Phys.}\ }\textbf {\bibinfo {volume} {43}},\ \bibinfo {pages}
  {357--459} (\bibinfo {year} {1994})}\BibitemShut {NoStop}%
\bibitem [{\citenamefont {Farkas}(1927)}]{Farkas+1927+236+242}%
  \BibitemOpen
  \bibfield  {author} {\bibinfo {author} {\bibfnamefont {L.}~\bibnamefont
  {Farkas}},\ }\bibfield  {title} {\enquote {\bibinfo {title}
  {Keimbildungsgeschwindigkeit in übersättigten dämpfen},}\ }\href@noop {}
  {\bibfield  {journal} {\bibinfo  {journal} {Zeitschrift für Physikalische
  Chemie}\ }\textbf {\bibinfo {volume} {125U}},\ \bibinfo {pages} {236--242}
  (\bibinfo {year} {1927})}\BibitemShut {NoStop}%
\bibitem [{\citenamefont {Volmer}\ and\ \citenamefont
  {Schultze}(1931)}]{VolmerSchultze+1931+1+22}%
  \BibitemOpen
  \bibfield  {author} {\bibinfo {author} {\bibfnamefont {M.}~\bibnamefont
  {Volmer}}\ and\ \bibinfo {author} {\bibfnamefont {W.}~\bibnamefont
  {Schultze}},\ }\bibfield  {title} {\enquote {\bibinfo {title} {Kondensation
  an kristallen},}\ }\href@noop {} {\bibfield  {journal} {\bibinfo  {journal}
  {Zeitschrift für Physikalische Chemie}\ }\textbf {\bibinfo {volume}
  {156A}},\ \bibinfo {pages} {1--22} (\bibinfo {year} {1931})}\BibitemShut
  {NoStop}%
\bibitem [{\citenamefont {Becker}\ and\ \citenamefont
  {Döring}(1935)}]{https://doi.org/10.1002/andp.19354160806}%
  \BibitemOpen
  \bibfield  {author} {\bibinfo {author} {\bibfnamefont {R.}~\bibnamefont
  {Becker}}\ and\ \bibinfo {author} {\bibfnamefont {W.}~\bibnamefont
  {Döring}},\ }\bibfield  {title} {\enquote {\bibinfo {title} {Kinetische
  behandlung der keimbildung in übersättigten dämpfen},}\ }\href@noop {}
  {\bibfield  {journal} {\bibinfo  {journal} {Annalen der Physik}\ }\textbf
  {\bibinfo {volume} {416}},\ \bibinfo {pages} {719--752} (\bibinfo {year}
  {1935})}\BibitemShut {NoStop}%
\bibitem [{\citenamefont {Zeldovich}(1943)}]{zeldovich1943theory}%
  \BibitemOpen
  \bibfield  {author} {\bibinfo {author} {\bibfnamefont {Y.~B.}\ \bibnamefont
  {Zeldovich}},\ }\bibfield  {title} {\enquote {\bibinfo {title} {On the theory
  of new phase formation: cavitation},}\ }\href@noop {} {\bibfield  {journal}
  {\bibinfo  {journal} {Acta Physicochem., USSR}\ }\textbf {\bibinfo {volume}
  {18}},\ \bibinfo {pages} {1} (\bibinfo {year} {1943})}\BibitemShut {NoStop}%
\bibitem [{\citenamefont {Frenkel}(1955)}]{Frenkel}%
  \BibitemOpen
  \bibfield  {author} {\bibinfo {author} {\bibfnamefont {J.}~\bibnamefont
  {Frenkel}},\ }\href@noop {} {\emph {\bibinfo {title} {Kinetic Theory of
  Liquids}}}\ (\bibinfo  {publisher} {Dover},\ \bibinfo {address} {New York},\
  \bibinfo {year} {1955})\BibitemShut {NoStop}%
\bibitem [{\citenamefont {Berry}, \citenamefont {Rice},\ and\ \citenamefont
  {Ross}(1980)}]{BRR}%
  \BibitemOpen
  \bibfield  {author} {\bibinfo {author} {\bibfnamefont {R.~S.}\ \bibnamefont
  {Berry}}, \bibinfo {author} {\bibfnamefont {S.~A.}\ \bibnamefont {Rice}}, \
  and\ \bibinfo {author} {\bibfnamefont {J.}~\bibnamefont {Ross}},\ }\href@noop
  {} {\emph {\bibinfo {title} {Physical Chemistry}}}\ (\bibinfo  {publisher}
  {John Wiley \& Sons},\ \bibinfo {address} {Hoboken, NJ},\ \bibinfo {year}
  {1980})\BibitemShut {NoStop}%
\bibitem [{\citenamefont {Silbey}, \citenamefont {Alberty},\ and\ \citenamefont
  {M.G.}(2004)}]{silbey2004physical}%
  \BibitemOpen
  \bibfield  {author} {\bibinfo {author} {\bibfnamefont {R.}~\bibnamefont
  {Silbey}}, \bibinfo {author} {\bibfnamefont {R.}~\bibnamefont {Alberty}}, \
  and\ \bibinfo {author} {\bibfnamefont {B.}~\bibnamefont {M.G.}},\ }\href@noop
  {} {\emph {\bibinfo {title} {Physical Chemistry}}}\ (\bibinfo  {publisher}
  {Wiley and Sons},\ \bibinfo {address} {Hoboken, NJ},\ \bibinfo {year}
  {2004})\BibitemShut {NoStop}%
\bibitem [{\citenamefont {Talanquer}(2005)}]{Talanquer2005phase}%
  \BibitemOpen
  \bibfield  {author} {\bibinfo {author} {\bibfnamefont {V.}~\bibnamefont
  {Talanquer}},\ }\bibfield  {title} {\enquote {\bibinfo {title} {Phase
  behavior of self-associating fluids with weaker dispersion interactions
  between bonded particles},}\ }\href@noop {} {\bibfield  {journal} {\bibinfo
  {journal} {J. Chem. Phys.}\ }\textbf {\bibinfo {volume} {122}},\ \bibinfo
  {pages} {154510} (\bibinfo {year} {2005})}\BibitemShut {NoStop}%
\bibitem [{\citenamefont {Wertheim}(1984{\natexlab{a}})}]{Wertheim1984fluidsI}%
  \BibitemOpen
  \bibfield  {author} {\bibinfo {author} {\bibfnamefont {M.~S.}\ \bibnamefont
  {Wertheim}},\ }\bibfield  {title} {\enquote {\bibinfo {title} {Fluids with
  highly directional attractive forces. i. statistical thermodynamics},}\
  }\href@noop {} {\bibfield  {journal} {\bibinfo  {journal} {J. Stat. Phys.}\
  }\textbf {\bibinfo {volume} {35}},\ \bibinfo {pages} {19--34} (\bibinfo
  {year} {1984}{\natexlab{a}})}\BibitemShut {NoStop}%
\bibitem [{\citenamefont
  {Wertheim}(1984{\natexlab{b}})}]{Wertheim1984fluidsII}%
  \BibitemOpen
  \bibfield  {author} {\bibinfo {author} {\bibfnamefont {M.~S.}\ \bibnamefont
  {Wertheim}},\ }\bibfield  {title} {\enquote {\bibinfo {title} {Fluids with
  highly directional attractive forces. ii. thermodynamic perturbation theory
  and integral equations},}\ }\href@noop {} {\bibfield  {journal} {\bibinfo
  {journal} {J. Stat. Phys.}\ }\textbf {\bibinfo {volume} {35}},\ \bibinfo
  {pages} {35--47} (\bibinfo {year} {1984}{\natexlab{b}})}\BibitemShut
  {NoStop}%
\bibitem [{\citenamefont
  {Wertheim}(1986{\natexlab{a}})}]{Wertheim1986fluidsIII}%
  \BibitemOpen
  \bibfield  {author} {\bibinfo {author} {\bibfnamefont {M.~S.}\ \bibnamefont
  {Wertheim}},\ }\bibfield  {title} {\enquote {\bibinfo {title} {Fluids with
  highly directional attractive forces. iii. multiple attraction sites},}\
  }\href@noop {} {\bibfield  {journal} {\bibinfo  {journal} {J. Stat. Phys.}\
  }\textbf {\bibinfo {volume} {42}},\ \bibinfo {pages} {459--476} (\bibinfo
  {year} {1986}{\natexlab{a}})}\BibitemShut {NoStop}%
\bibitem [{\citenamefont
  {Wertheim}(1986{\natexlab{b}})}]{Wertheim1986fluidsIV}%
  \BibitemOpen
  \bibfield  {author} {\bibinfo {author} {\bibfnamefont {M.~S.}\ \bibnamefont
  {Wertheim}},\ }\bibfield  {title} {\enquote {\bibinfo {title} {Fluids with
  highly directional attractive forces. iv. equilibrium polymerization},}\
  }\href@noop {} {\bibfield  {journal} {\bibinfo  {journal} {J. Stat. Phys.}\
  }\textbf {\bibinfo {volume} {42}},\ \bibinfo {pages} {477--492} (\bibinfo
  {year} {1986}{\natexlab{b}})}\BibitemShut {NoStop}%
\bibitem [{\citenamefont {Chapman}\ \emph {et~al.}(1986)\citenamefont
  {Chapman}, \citenamefont {Gubbins}, \citenamefont {Joslin},\ and\
  \citenamefont {Gray}}]{Chapman1986theory}%
  \BibitemOpen
  \bibfield  {author} {\bibinfo {author} {\bibfnamefont {W.~G.}\ \bibnamefont
  {Chapman}}, \bibinfo {author} {\bibfnamefont {K.}~\bibnamefont {Gubbins}},
  \bibinfo {author} {\bibfnamefont {C.}~\bibnamefont {Joslin}}, \ and\ \bibinfo
  {author} {\bibfnamefont {C.}~\bibnamefont {Gray}},\ }\bibfield  {title}
  {\enquote {\bibinfo {title} {Theory and simulation of associating liquid
  mixtures},}\ }\href@noop {} {\bibfield  {journal} {\bibinfo  {journal} {Fluid
  Phase Equilibria}\ }\textbf {\bibinfo {volume} {29}},\ \bibinfo {pages}
  {337--346} (\bibinfo {year} {1986})}\BibitemShut {NoStop}%
\bibitem [{\citenamefont {Joslin}\ \emph {et~al.}(1987)\citenamefont {Joslin},
  \citenamefont {Gray}, \citenamefont {Chapman},\ and\ \citenamefont
  {Gubbins}}]{Joslin1987theory}%
  \BibitemOpen
  \bibfield  {author} {\bibinfo {author} {\bibfnamefont {C.}~\bibnamefont
  {Joslin}}, \bibinfo {author} {\bibfnamefont {C.}~\bibnamefont {Gray}},
  \bibinfo {author} {\bibfnamefont {W.}~\bibnamefont {Chapman}}, \ and\
  \bibinfo {author} {\bibfnamefont {K.}~\bibnamefont {Gubbins}},\ }\bibfield
  {title} {\enquote {\bibinfo {title} {Theory and simulation of associating
  liquid mixtures. ii},}\ }\href@noop {} {\bibfield  {journal} {\bibinfo
  {journal} {Molecular Physics}\ }\textbf {\bibinfo {volume} {62}},\ \bibinfo
  {pages} {843--860} (\bibinfo {year} {1987})}\BibitemShut {NoStop}%
\bibitem [{\citenamefont {Jackson}, \citenamefont {Chapman},\ and\
  \citenamefont {Gubbins}(1988)}]{JCG1988AL}%
  \BibitemOpen
  \bibfield  {author} {\bibinfo {author} {\bibfnamefont {G.}~\bibnamefont
  {Jackson}}, \bibinfo {author} {\bibfnamefont {W.~G.}\ \bibnamefont
  {Chapman}}, \ and\ \bibinfo {author} {\bibfnamefont {K.~E.}\ \bibnamefont
  {Gubbins}},\ }\bibfield  {title} {\enquote {\bibinfo {title} {Phase
  equilibria of associating fluids},}\ }\href@noop {} {\bibfield  {journal}
  {\bibinfo  {journal} {Molecular Physics}\ }\textbf {\bibinfo {volume} {65}},\
  \bibinfo {pages} {1--31} (\bibinfo {year} {1988})}\BibitemShut {NoStop}%
\bibitem [{\citenamefont {Chapman}, \citenamefont {Jackson},\ and\
  \citenamefont {Gubbins}(1988)}]{CJG1988AL}%
  \BibitemOpen
  \bibfield  {author} {\bibinfo {author} {\bibfnamefont {W.~G.}\ \bibnamefont
  {Chapman}}, \bibinfo {author} {\bibfnamefont {G.}~\bibnamefont {Jackson}}, \
  and\ \bibinfo {author} {\bibfnamefont {K.~E.}\ \bibnamefont {Gubbins}},\
  }\bibfield  {title} {\enquote {\bibinfo {title} {Phase equilibria of
  associating fluids},}\ }\href@noop {} {\bibfield  {journal} {\bibinfo
  {journal} {Molecular Physics}\ }\textbf {\bibinfo {volume} {65}},\ \bibinfo
  {pages} {1057--1079} (\bibinfo {year} {1988})}\BibitemShut {NoStop}%
\bibitem [{\citenamefont {Chapman}\ \emph {et~al.}(1989)\citenamefont
  {Chapman}, \citenamefont {Gubbins}, \citenamefont {Jackson},\ and\
  \citenamefont {Radosz}}]{CHAPMAN1989}%
  \BibitemOpen
  \bibfield  {author} {\bibinfo {author} {\bibfnamefont {W.}~\bibnamefont
  {Chapman}}, \bibinfo {author} {\bibfnamefont {K.}~\bibnamefont {Gubbins}},
  \bibinfo {author} {\bibfnamefont {G.}~\bibnamefont {Jackson}}, \ and\
  \bibinfo {author} {\bibfnamefont {M.}~\bibnamefont {Radosz}},\ }\bibfield
  {title} {\enquote {\bibinfo {title} {Saft: Equation-of-state solution model
  for associating fluids},}\ }\href {\doibase
  https://doi.org/10.1016/0378-3812(89)80308-5} {\bibfield  {journal} {\bibinfo
   {journal} {Fluid Phase Equilibria}\ }\textbf {\bibinfo {volume} {52}},\
  \bibinfo {pages} {31--38} (\bibinfo {year} {1989})}\BibitemShut {NoStop}%
\bibitem [{\citenamefont {M{\"u}ller}\ and\ \citenamefont
  {Gubbins}(2001)}]{Muller2001saft}%
  \BibitemOpen
  \bibfield  {author} {\bibinfo {author} {\bibfnamefont {E.~A.}\ \bibnamefont
  {M{\"u}ller}}\ and\ \bibinfo {author} {\bibfnamefont {K.~E.}\ \bibnamefont
  {Gubbins}},\ }\bibfield  {title} {\enquote {\bibinfo {title} {Molecular-based
  equations of state for associating fluids: A review of saft and related
  approaches},}\ }\href@noop {} {\bibfield  {journal} {\bibinfo  {journal}
  {Industrial \& engineering chemistry research}\ }\textbf {\bibinfo {volume}
  {40}},\ \bibinfo {pages} {2193--2211} (\bibinfo {year} {2001})}\BibitemShut
  {NoStop}%
\bibitem [{\citenamefont {Fouad}\ \emph {et~al.}(2016)\citenamefont {Fouad},
  \citenamefont {Haghmoradi}, \citenamefont {Wang}, \citenamefont {Bansal},
  \citenamefont {{Al Hammadi}}, \citenamefont {Asthagiri}, \citenamefont
  {Djamali}, \citenamefont {Cox},\ and\ \citenamefont {Chapman}}]{FOUAD201662}%
  \BibitemOpen
  \bibfield  {author} {\bibinfo {author} {\bibfnamefont {W.~A.}\ \bibnamefont
  {Fouad}}, \bibinfo {author} {\bibfnamefont {A.}~\bibnamefont {Haghmoradi}},
  \bibinfo {author} {\bibfnamefont {L.}~\bibnamefont {Wang}}, \bibinfo {author}
  {\bibfnamefont {A.}~\bibnamefont {Bansal}}, \bibinfo {author} {\bibfnamefont
  {A.}~\bibnamefont {{Al Hammadi}}}, \bibinfo {author} {\bibfnamefont
  {D.}~\bibnamefont {Asthagiri}}, \bibinfo {author} {\bibfnamefont
  {E.}~\bibnamefont {Djamali}}, \bibinfo {author} {\bibfnamefont {K.~R.}\
  \bibnamefont {Cox}}, \ and\ \bibinfo {author} {\bibfnamefont {W.~G.}\
  \bibnamefont {Chapman}},\ }\bibfield  {title} {\enquote {\bibinfo {title}
  {Extensions of the saft model for complex association in the bulk and
  interface},}\ }\href@noop {} {\bibfield  {journal} {\bibinfo  {journal}
  {Fluid Phase Equilibria}\ }\textbf {\bibinfo {volume} {416}},\ \bibinfo
  {pages} {62--71} (\bibinfo {year} {2016})}\BibitemShut {NoStop}%
\bibitem [{\citenamefont {Chapman}(1990)}]{10.1063/1.458711}%
  \BibitemOpen
  \bibfield  {author} {\bibinfo {author} {\bibfnamefont {W.~G.}\ \bibnamefont
  {Chapman}},\ }\bibfield  {title} {\enquote {\bibinfo {title} {Prediction of
  the thermodynamic properties of associating lennard‐jones fluids: Theory
  and simulation},}\ }\href@noop {} {\bibfield  {journal} {\bibinfo  {journal}
  {J. Chem. Phys.}\ }\textbf {\bibinfo {volume} {93}},\ \bibinfo {pages}
  {4299--4304} (\bibinfo {year} {1990})}\BibitemShut {NoStop}%
\bibitem [{\citenamefont {Marshall}(2025)}]{MARSHALL2025114461}%
  \BibitemOpen
  \bibfield  {author} {\bibinfo {author} {\bibfnamefont {B.~D.}\ \bibnamefont
  {Marshall}},\ }\bibfield  {title} {\enquote {\bibinfo {title} {Thermodynamic
  perturbation theory for associating fluids with coupling to isotropic
  attractions},}\ }\href@noop {} {\bibfield  {journal} {\bibinfo  {journal}
  {Fluid Phase Equilibria}\ }\textbf {\bibinfo {volume} {597}},\ \bibinfo
  {pages} {114461} (\bibinfo {year} {2025})}\BibitemShut {NoStop}%
\bibitem [{\citenamefont {Pappu}\ \emph {et~al.}(2023)\citenamefont {Pappu},
  \citenamefont {Cohen}, \citenamefont {Dar}, \citenamefont {Farag},\ and\
  \citenamefont {Kar}}]{doi:10.1021/acs.chemrev.2c00814}%
  \BibitemOpen
  \bibfield  {author} {\bibinfo {author} {\bibfnamefont {R.~V.}\ \bibnamefont
  {Pappu}}, \bibinfo {author} {\bibfnamefont {S.~R.}\ \bibnamefont {Cohen}},
  \bibinfo {author} {\bibfnamefont {F.}~\bibnamefont {Dar}}, \bibinfo {author}
  {\bibfnamefont {M.}~\bibnamefont {Farag}}, \ and\ \bibinfo {author}
  {\bibfnamefont {M.}~\bibnamefont {Kar}},\ }\bibfield  {title} {\enquote
  {\bibinfo {title} {Phase transitions of associative biomacromolecules},}\
  }\href@noop {} {\bibfield  {journal} {\bibinfo  {journal} {Chemical Reviews}\
  }\textbf {\bibinfo {volume} {123}},\ \bibinfo {pages} {8945--8987} (\bibinfo
  {year} {2023})}\BibitemShut {NoStop}%
\bibitem [{\citenamefont {He}\ and\ \citenamefont
  {Lubchenko}(2025)}]{HLKnowledge}%
  \BibitemOpen
  \bibfield  {author} {\bibinfo {author} {\bibfnamefont {Y.}~\bibnamefont
  {He}}\ and\ \bibinfo {author} {\bibfnamefont {V.}~\bibnamefont {Lubchenko}},\
  }\bibfield  {title} {\enquote {\bibinfo {title} {Knowledge as a breaking of
  ergodicity},}\ }\href@noop {} {\bibfield  {journal} {\bibinfo  {journal}
  {Neural Comput.}\ }\textbf {\bibinfo {volume} {37}},\ \bibinfo {pages}
  {742--792} (\bibinfo {year} {2025})}\BibitemShut {NoStop}%
\bibitem [{\citenamefont {Chan}\ and\ \citenamefont {Lubchenko}(2015)}]{CL_LG}%
  \BibitemOpen
  \bibfield  {author} {\bibinfo {author} {\bibfnamefont {H.~Y.}\ \bibnamefont
  {Chan}}\ and\ \bibinfo {author} {\bibfnamefont {V.}~\bibnamefont
  {Lubchenko}},\ }\bibfield  {title} {\enquote {\bibinfo {title} {{Pressure in
  the Landau-Ginzburg functional: Pascal's law, nucleation in fluid mixtures, a
  meanfield theory of amphiphilic action, and interface wetting in glassy
  liquids}},}\ }\href@noop {} {\bibfield  {journal} {\bibinfo  {journal} {J.
  Chem. Phys.}\ }\textbf {\bibinfo {volume} {143}},\ \bibinfo {pages} {124502}
  (\bibinfo {year} {2015})}\BibitemShut {NoStop}%
\bibitem [{\citenamefont {Lubchenko}(2020)}]{LSurvey}%
  \BibitemOpen
  \bibfield  {author} {\bibinfo {author} {\bibfnamefont {V.}~\bibnamefont
  {Lubchenko}},\ }\href@noop {} {\emph {\bibinfo {title} {{Basic Notions of
  Thermodynamics and Quantum Mechanics for Natural Sciences}}}}\ (\bibinfo
  {publisher} {University of Houston},\ \bibinfo {year} {2020})\ \bibinfo
  {note} {https://uhlibraries.pressbooks.pub/lubchenkosurveypchem/}\BibitemShut
  {NoStop}%
\bibitem [{\citenamefont {Landau}\ and\ \citenamefont
  {Lifshitz}(1980)}]{LLstat}%
  \BibitemOpen
  \bibfield  {author} {\bibinfo {author} {\bibfnamefont {L.~D.}\ \bibnamefont
  {Landau}}\ and\ \bibinfo {author} {\bibfnamefont {E.~M.}\ \bibnamefont
  {Lifshitz}},\ }\href@noop {} {\emph {\bibinfo {title} {Statistical
  Mechanics}}}\ (\bibinfo  {publisher} {Pergamon Press},\ \bibinfo {address}
  {New York},\ \bibinfo {year} {1980})\BibitemShut {NoStop}%
\bibitem [{\citenamefont {Huang}\ \emph {et~al.}(2021)\citenamefont {Huang},
  \citenamefont {Mahrt}, \citenamefont {Xu}, \citenamefont {Shiraiwa},
  \citenamefont {Zuend},\ and\ \citenamefont
  {Bertram}}]{doi:10.1073/pnas.2102512118}%
  \BibitemOpen
  \bibfield  {author} {\bibinfo {author} {\bibfnamefont {Y.}~\bibnamefont
  {Huang}}, \bibinfo {author} {\bibfnamefont {F.}~\bibnamefont {Mahrt}},
  \bibinfo {author} {\bibfnamefont {S.}~\bibnamefont {Xu}}, \bibinfo {author}
  {\bibfnamefont {M.}~\bibnamefont {Shiraiwa}}, \bibinfo {author}
  {\bibfnamefont {A.}~\bibnamefont {Zuend}}, \ and\ \bibinfo {author}
  {\bibfnamefont {A.~K.}\ \bibnamefont {Bertram}},\ }\bibfield  {title}
  {\enquote {\bibinfo {title} {Coexistence of three liquid phases in individual
  atmospheric aerosol particles},}\ }\href@noop {} {\bibfield  {journal}
  {\bibinfo  {journal} {Proc. Natl. Acad. Sci. U.~S.~A.}\ }\textbf {\bibinfo
  {volume} {118}},\ \bibinfo {pages} {e2102512118} (\bibinfo {year}
  {2021})}\BibitemShut {NoStop}%
\bibitem [{\citenamefont {Lubchenko}(2015)}]{L_AP}%
  \BibitemOpen
  \bibfield  {author} {\bibinfo {author} {\bibfnamefont {V.}~\bibnamefont
  {Lubchenko}},\ }\bibfield  {title} {\enquote {\bibinfo {title} {Theory of the
  structural glass transition: A pedagogical review},}\ }\href@noop {}
  {\bibfield  {journal} {\bibinfo  {journal} {Adv. Phys.}\ }\textbf {\bibinfo
  {volume} {64}},\ \bibinfo {pages} {283--443} (\bibinfo {year}
  {2015})}\BibitemShut {NoStop}%
\bibitem [{\citenamefont {Vorontsova}, \citenamefont {Maes},\ and\
  \citenamefont {Vekilov}(2015)}]{C4FD00217B}%
  \BibitemOpen
  \bibfield  {author} {\bibinfo {author} {\bibfnamefont {M.~A.}\ \bibnamefont
  {Vorontsova}}, \bibinfo {author} {\bibfnamefont {D.}~\bibnamefont {Maes}}, \
  and\ \bibinfo {author} {\bibfnamefont {P.~G.}\ \bibnamefont {Vekilov}},\
  }\bibfield  {title} {\enquote {\bibinfo {title} {Recent advances in the
  understanding of two-step nucleation of protein crystals},}\ }\href@noop {}
  {\bibfield  {journal} {\bibinfo  {journal} {Faraday Discuss.}\ }\textbf
  {\bibinfo {volume} {179}},\ \bibinfo {pages} {27--40} (\bibinfo {year}
  {2015})}\BibitemShut {NoStop}%
\bibitem [{\citenamefont {Kaptay}(2024)}]{ma17246048}%
  \BibitemOpen
  \bibfield  {author} {\bibinfo {author} {\bibfnamefont {G.}~\bibnamefont
  {Kaptay}},\ }\bibfield  {title} {\enquote {\bibinfo {title} {The generalized
  phase rule, the extended definition of the degree of freedom, the component
  rule and the seven independent non-compositional state variables: To the
  150th anniversary of the phase rule of gibbs},}\ }\href@noop {} {\bibfield
  {journal} {\bibinfo  {journal} {Materials}\ }\textbf {\bibinfo {volume} {17}}
  (\bibinfo {year} {2024})}\BibitemShut {NoStop}%
\bibitem [{\citenamefont {Opdam}\ \emph {et~al.}(2023)\citenamefont {Opdam},
  \citenamefont {Peters}, \citenamefont {Wensink},\ and\ \citenamefont
  {Tuinier}}]{doi:10.1021/acs.jpclett.2c03138}%
  \BibitemOpen
  \bibfield  {author} {\bibinfo {author} {\bibfnamefont {J.}~\bibnamefont
  {Opdam}}, \bibinfo {author} {\bibfnamefont {V.~F.~D.}\ \bibnamefont
  {Peters}}, \bibinfo {author} {\bibfnamefont {H.~H.}\ \bibnamefont {Wensink}},
  \ and\ \bibinfo {author} {\bibfnamefont {R.}~\bibnamefont {Tuinier}},\
  }\bibfield  {title} {\enquote {\bibinfo {title} {Multiphase coexistence in
  binary hard colloidal mixtures: Predictions from a simple algebraic
  theory},}\ }\href@noop {} {\bibfield  {journal} {\bibinfo  {journal} {The
  Journal of Physical Chemistry Letters}\ }\textbf {\bibinfo {volume} {14}},\
  \bibinfo {pages} {199--206} (\bibinfo {year} {2023})}\BibitemShut {NoStop}%
\bibitem [{\citenamefont {Wu}\ \emph {et~al.}(2025)\citenamefont {Wu},
  \citenamefont {King}, \citenamefont {Qiu}, \citenamefont {Farag},
  \citenamefont {Pappu},\ and\ \citenamefont {Lew}}]{Wu2025}%
  \BibitemOpen
  \bibfield  {author} {\bibinfo {author} {\bibfnamefont {T.}~\bibnamefont
  {Wu}}, \bibinfo {author} {\bibfnamefont {M.~R.}\ \bibnamefont {King}},
  \bibinfo {author} {\bibfnamefont {Y.}~\bibnamefont {Qiu}}, \bibinfo {author}
  {\bibfnamefont {M.}~\bibnamefont {Farag}}, \bibinfo {author} {\bibfnamefont
  {R.~V.}\ \bibnamefont {Pappu}}, \ and\ \bibinfo {author} {\bibfnamefont
  {M.~D.}\ \bibnamefont {Lew}},\ }\bibfield  {title} {\enquote {\bibinfo
  {title} {Single-fluorogen imaging reveals distinct environmental and
  structural features of biomolecular condensates},}\ }\href@noop {} {\bibfield
   {journal} {\bibinfo  {journal} {Nature Physics}\ }\textbf {\bibinfo {volume}
  {21}},\ \bibinfo {pages} {778--786} (\bibinfo {year} {2025})}\BibitemShut
  {NoStop}%
\bibitem [{\citenamefont {Lan}\ \emph {et~al.}(2023)\citenamefont {Lan},
  \citenamefont {Kim}, \citenamefont {Ulferts}, \citenamefont {Aprile-Garcia},
  \citenamefont {Weyrauch}, \citenamefont {Anandamurugan}, \citenamefont
  {Grosse}, \citenamefont {Sawarkar}, \citenamefont {Reinhardt},\ and\
  \citenamefont {Hugel}}]{Lan2023}%
  \BibitemOpen
  \bibfield  {author} {\bibinfo {author} {\bibfnamefont {C.}~\bibnamefont
  {Lan}}, \bibinfo {author} {\bibfnamefont {J.}~\bibnamefont {Kim}}, \bibinfo
  {author} {\bibfnamefont {S.}~\bibnamefont {Ulferts}}, \bibinfo {author}
  {\bibfnamefont {F.}~\bibnamefont {Aprile-Garcia}}, \bibinfo {author}
  {\bibfnamefont {S.}~\bibnamefont {Weyrauch}}, \bibinfo {author}
  {\bibfnamefont {A.}~\bibnamefont {Anandamurugan}}, \bibinfo {author}
  {\bibfnamefont {R.}~\bibnamefont {Grosse}}, \bibinfo {author} {\bibfnamefont
  {R.}~\bibnamefont {Sawarkar}}, \bibinfo {author} {\bibfnamefont
  {A.}~\bibnamefont {Reinhardt}}, \ and\ \bibinfo {author} {\bibfnamefont
  {T.}~\bibnamefont {Hugel}},\ }\bibfield  {title} {\enquote {\bibinfo {title}
  {Quantitative real-time in-cell imaging reveals heterogeneous clusters of
  proteins prior to condensation},}\ }\href@noop {} {\bibfield  {journal}
  {\bibinfo  {journal} {Nature Communications}\ }\textbf {\bibinfo {volume}
  {14}},\ \bibinfo {pages} {4831} (\bibinfo {year} {2023})}\BibitemShut
  {NoStop}%
\bibitem [{\citenamefont {Kar}\ \emph {et~al.}(2024)\citenamefont {Kar},
  \citenamefont {Vogel}, \citenamefont {Chauhan}, \citenamefont {Felekyan},
  \citenamefont {Ausserw{\"o}ger}, \citenamefont {Welsh}, \citenamefont {Dar},
  \citenamefont {Kamath}, \citenamefont {Knowles}, \citenamefont {Hyman},
  \citenamefont {Seidel},\ and\ \citenamefont {Pappu}}]{Kar2024}%
  \BibitemOpen
  \bibfield  {author} {\bibinfo {author} {\bibfnamefont {M.}~\bibnamefont
  {Kar}}, \bibinfo {author} {\bibfnamefont {L.~T.}\ \bibnamefont {Vogel}},
  \bibinfo {author} {\bibfnamefont {G.}~\bibnamefont {Chauhan}}, \bibinfo
  {author} {\bibfnamefont {S.}~\bibnamefont {Felekyan}}, \bibinfo {author}
  {\bibfnamefont {H.}~\bibnamefont {Ausserw{\"o}ger}}, \bibinfo {author}
  {\bibfnamefont {T.~J.}\ \bibnamefont {Welsh}}, \bibinfo {author}
  {\bibfnamefont {F.}~\bibnamefont {Dar}}, \bibinfo {author} {\bibfnamefont
  {A.~R.}\ \bibnamefont {Kamath}}, \bibinfo {author} {\bibfnamefont {T.~P.~J.}\
  \bibnamefont {Knowles}}, \bibinfo {author} {\bibfnamefont {A.~A.}\
  \bibnamefont {Hyman}}, \bibinfo {author} {\bibfnamefont {C.~A.~M.}\
  \bibnamefont {Seidel}}, \ and\ \bibinfo {author} {\bibfnamefont {R.~V.}\
  \bibnamefont {Pappu}},\ }\bibfield  {title} {\enquote {\bibinfo {title}
  {Solutes unmask differences in clustering versus phase separation of fet
  proteins},}\ }\href@noop {} {\bibfield  {journal} {\bibinfo  {journal}
  {Nature Communications}\ }\textbf {\bibinfo {volume} {15}},\ \bibinfo {pages}
  {4408} (\bibinfo {year} {2024})}\BibitemShut {NoStop}%
\bibitem [{\citenamefont {Yanas}\ \emph {et~al.}(2024)\citenamefont {Yanas},
  \citenamefont {Shweta}, \citenamefont {Owens}, \citenamefont {Liu},\ and\
  \citenamefont {Goldman}}]{Yanas2024}%
  \BibitemOpen
  \bibfield  {author} {\bibinfo {author} {\bibfnamefont {A.}~\bibnamefont
  {Yanas}}, \bibinfo {author} {\bibfnamefont {H.}~\bibnamefont {Shweta}},
  \bibinfo {author} {\bibfnamefont {M.~C.}\ \bibnamefont {Owens}}, \bibinfo
  {author} {\bibfnamefont {K.~F.}\ \bibnamefont {Liu}}, \ and\ \bibinfo
  {author} {\bibfnamefont {Y.~E.}\ \bibnamefont {Goldman}},\ }\bibfield
  {title} {\enquote {\bibinfo {title} {Rna helicases ddx3x and ddx3y form
  nanometer-scale rna-protein clusters that support catalytic activity},}\
  }\href@noop {} {\bibfield  {journal} {\bibinfo  {journal} {Current Biology}\
  }\textbf {\bibinfo {volume} {34}},\ \bibinfo {pages} {5714--5727.e6}
  (\bibinfo {year} {2024})}\BibitemShut {NoStop}%
\bibitem [{\citenamefont {Hoffmann}\ \emph {et~al.}(2025)\citenamefont
  {Hoffmann}, \citenamefont {Ruff}, \citenamefont {Edu}, \citenamefont {Shinn},
  \citenamefont {Tromm}, \citenamefont {King}, \citenamefont {Pant},
  \citenamefont {Ausserwöger}, \citenamefont {Morgan}, \citenamefont
  {Knowles}, \citenamefont {Pappu},\ and\ \citenamefont
  {Milovanovic}}]{HOFFMANN2025168987}%
  \BibitemOpen
  \bibfield  {author} {\bibinfo {author} {\bibfnamefont {C.}~\bibnamefont
  {Hoffmann}}, \bibinfo {author} {\bibfnamefont {K.~M.}\ \bibnamefont {Ruff}},
  \bibinfo {author} {\bibfnamefont {I.~A.}\ \bibnamefont {Edu}}, \bibinfo
  {author} {\bibfnamefont {M.~K.}\ \bibnamefont {Shinn}}, \bibinfo {author}
  {\bibfnamefont {J.~V.}\ \bibnamefont {Tromm}}, \bibinfo {author}
  {\bibfnamefont {M.~R.}\ \bibnamefont {King}}, \bibinfo {author}
  {\bibfnamefont {A.}~\bibnamefont {Pant}}, \bibinfo {author} {\bibfnamefont
  {H.}~\bibnamefont {Ausserwöger}}, \bibinfo {author} {\bibfnamefont {J.~R.}\
  \bibnamefont {Morgan}}, \bibinfo {author} {\bibfnamefont {T.~P.}\
  \bibnamefont {Knowles}}, \bibinfo {author} {\bibfnamefont {R.~V.}\
  \bibnamefont {Pappu}}, \ and\ \bibinfo {author} {\bibfnamefont
  {D.}~\bibnamefont {Milovanovic}},\ }\bibfield  {title} {\enquote {\bibinfo
  {title} {Synapsin condensation is governed by sequence-encoded molecular
  grammars},}\ }\href@noop {} {\bibfield  {journal} {\bibinfo  {journal}
  {Journal of Molecular Biology}\ }\textbf {\bibinfo {volume} {437}},\ \bibinfo
  {pages} {168987} (\bibinfo {year} {2025})}\BibitemShut {NoStop}%
\bibitem [{\citenamefont {Sanchez~de Groot}\ \emph {et~al.}(2012)\citenamefont
  {Sanchez~de Groot}, \citenamefont {Torrent}, \citenamefont {Villar-Piqu\'e},
  \citenamefont {Lang}, \citenamefont {Ventura}, \citenamefont {Gsponer},\ and\
  \citenamefont {Babu}}]{10.1042/BST20120160}%
  \BibitemOpen
  \bibfield  {author} {\bibinfo {author} {\bibfnamefont {N.}~\bibnamefont
  {Sanchez~de Groot}}, \bibinfo {author} {\bibfnamefont {M.}~\bibnamefont
  {Torrent}}, \bibinfo {author} {\bibfnamefont {A.}~\bibnamefont
  {Villar-Piqu\'e}}, \bibinfo {author} {\bibfnamefont {B.}~\bibnamefont
  {Lang}}, \bibinfo {author} {\bibfnamefont {S.}~\bibnamefont {Ventura}},
  \bibinfo {author} {\bibfnamefont {J.}~\bibnamefont {Gsponer}}, \ and\
  \bibinfo {author} {\bibfnamefont {M.~M.}\ \bibnamefont {Babu}},\ }\bibfield
  {title} {\enquote {\bibinfo {title} {Evolutionary selection for protein
  aggregation},}\ }\href@noop {} {\bibfield  {journal} {\bibinfo  {journal}
  {Biochem. Soc. Trans.}\ }\textbf {\bibinfo {volume} {40}},\ \bibinfo {pages}
  {1032--1037} (\bibinfo {year} {2012})}\BibitemShut {NoStop}%
\bibitem [{\citenamefont {Monsellier}\ and\ \citenamefont
  {Chiti}(2007)}]{https://doi.org/10.1038/sj.embor.7401034}%
  \BibitemOpen
  \bibfield  {author} {\bibinfo {author} {\bibfnamefont {E.}~\bibnamefont
  {Monsellier}}\ and\ \bibinfo {author} {\bibfnamefont {F.}~\bibnamefont
  {Chiti}},\ }\bibfield  {title} {\enquote {\bibinfo {title} {Prevention of
  amyloid‐like aggregation as a driving force of protein evolution},}\
  }\href@noop {} {\bibfield  {journal} {\bibinfo  {journal} {EMBO reports}\
  }\textbf {\bibinfo {volume} {8}},\ \bibinfo {pages} {737--742} (\bibinfo
  {year} {2007})}\BibitemShut {NoStop}%
\bibitem [{\citenamefont {Weinreich}\ \emph {et~al.}(2006)\citenamefont
  {Weinreich}, \citenamefont {Delaney}, \citenamefont {DePristo},\ and\
  \citenamefont {Hartl}}]{doi:10.1126/science.1123539}%
  \BibitemOpen
  \bibfield  {author} {\bibinfo {author} {\bibfnamefont {D.~M.}\ \bibnamefont
  {Weinreich}}, \bibinfo {author} {\bibfnamefont {N.~F.}\ \bibnamefont
  {Delaney}}, \bibinfo {author} {\bibfnamefont {M.~A.}\ \bibnamefont
  {DePristo}}, \ and\ \bibinfo {author} {\bibfnamefont {D.~L.}\ \bibnamefont
  {Hartl}},\ }\bibfield  {title} {\enquote {\bibinfo {title} {Darwinian
  evolution can follow only very few mutational paths to fitter proteins},}\
  }\href@noop {} {\bibfield  {journal} {\bibinfo  {journal} {Science}\ }\textbf
  {\bibinfo {volume} {312}},\ \bibinfo {pages} {111--114} (\bibinfo {year}
  {2006})}\BibitemShut {NoStop}%
\bibitem [{\citenamefont {Keyport~Kik}\ \emph {et~al.}(2024)\citenamefont
  {Keyport~Kik}, \citenamefont {Christopher}, \citenamefont {Glauninger},
  \citenamefont {Hickernell}, \citenamefont {Bard}, \citenamefont {Lin},
  \citenamefont {Squires}, \citenamefont {Ford}, \citenamefont {Sosnick},\ and\
  \citenamefont {Drummond}}]{KeyportKik2024}%
  \BibitemOpen
  \bibfield  {author} {\bibinfo {author} {\bibfnamefont {S.}~\bibnamefont
  {Keyport~Kik}}, \bibinfo {author} {\bibfnamefont {D.}~\bibnamefont
  {Christopher}}, \bibinfo {author} {\bibfnamefont {H.}~\bibnamefont
  {Glauninger}}, \bibinfo {author} {\bibfnamefont {C.~W.}\ \bibnamefont
  {Hickernell}}, \bibinfo {author} {\bibfnamefont {J.~A.~M.}\ \bibnamefont
  {Bard}}, \bibinfo {author} {\bibfnamefont {K.~M.}\ \bibnamefont {Lin}},
  \bibinfo {author} {\bibfnamefont {A.~H.}\ \bibnamefont {Squires}}, \bibinfo
  {author} {\bibfnamefont {M.}~\bibnamefont {Ford}}, \bibinfo {author}
  {\bibfnamefont {T.~R.}\ \bibnamefont {Sosnick}}, \ and\ \bibinfo {author}
  {\bibfnamefont {D.~A.}\ \bibnamefont {Drummond}},\ }\bibfield  {title}
  {\enquote {\bibinfo {title} {An adaptive biomolecular condensation response
  is conserved across environmentally divergent species},}\ }\href@noop {}
  {\bibfield  {journal} {\bibinfo  {journal} {Nature Comm.}\ }\textbf {\bibinfo
  {volume} {15}},\ \bibinfo {pages} {3127} (\bibinfo {year}
  {2024})}\BibitemShut {NoStop}%
\bibitem [{\citenamefont {Patil}\ \emph {et~al.}(2023)\citenamefont {Patil},
  \citenamefont {Strom}, \citenamefont {Paulo}, \citenamefont {Collings},
  \citenamefont {Ruff}, \citenamefont {Shinn}, \citenamefont {Sankar},
  \citenamefont {Cervantes}, \citenamefont {Wauer}, \citenamefont
  {St.~Laurent}, \citenamefont {Xu}, \citenamefont {Becker}, \citenamefont
  {Gygi}, \citenamefont {Pappu}, \citenamefont {Brangwynne},\ and\
  \citenamefont {Kadoch}}]{Patil2023}%
  \BibitemOpen
  \bibfield  {author} {\bibinfo {author} {\bibfnamefont {A.}~\bibnamefont
  {Patil}}, \bibinfo {author} {\bibfnamefont {A.~R.}\ \bibnamefont {Strom}},
  \bibinfo {author} {\bibfnamefont {J.~A.}\ \bibnamefont {Paulo}}, \bibinfo
  {author} {\bibfnamefont {C.~K.}\ \bibnamefont {Collings}}, \bibinfo {author}
  {\bibfnamefont {K.~M.}\ \bibnamefont {Ruff}}, \bibinfo {author}
  {\bibfnamefont {M.~K.}\ \bibnamefont {Shinn}}, \bibinfo {author}
  {\bibfnamefont {A.}~\bibnamefont {Sankar}}, \bibinfo {author} {\bibfnamefont
  {K.~S.}\ \bibnamefont {Cervantes}}, \bibinfo {author} {\bibfnamefont
  {T.}~\bibnamefont {Wauer}}, \bibinfo {author} {\bibfnamefont {J.~D.}\
  \bibnamefont {St.~Laurent}}, \bibinfo {author} {\bibfnamefont
  {G.}~\bibnamefont {Xu}}, \bibinfo {author} {\bibfnamefont {L.~A.}\
  \bibnamefont {Becker}}, \bibinfo {author} {\bibfnamefont {S.~P.}\
  \bibnamefont {Gygi}}, \bibinfo {author} {\bibfnamefont {R.~V.}\ \bibnamefont
  {Pappu}}, \bibinfo {author} {\bibfnamefont {C.~P.}\ \bibnamefont
  {Brangwynne}}, \ and\ \bibinfo {author} {\bibfnamefont {C.}~\bibnamefont
  {Kadoch}},\ }\bibfield  {title} {\enquote {\bibinfo {title} {A disordered
  region controls cbaf activity via condensation and partner recruitment},}\
  }\href@noop {} {\bibfield  {journal} {\bibinfo  {journal} {Cell}\ }\textbf
  {\bibinfo {volume} {186}},\ \bibinfo {pages} {4936--4955.e26} (\bibinfo
  {year} {2023})}\BibitemShut {NoStop}%
\bibitem [{\citenamefont {Biswas}\ and\ \citenamefont
  {Potoyan}(2025)}]{10.1371/journal.pcbi.1012826}%
  \BibitemOpen
  \bibfield  {author} {\bibinfo {author} {\bibfnamefont {S.}~\bibnamefont
  {Biswas}}\ and\ \bibinfo {author} {\bibfnamefont {D.~A.}\ \bibnamefont
  {Potoyan}},\ }\bibfield  {title} {\enquote {\bibinfo {title} {Decoding
  biomolecular condensate dynamics: an energy landscape approach},}\ }\href
  {\doibase 10.1371/journal.pcbi.1012826} {\bibfield  {journal} {\bibinfo
  {journal} {PLOS Computational Biology}\ }\textbf {\bibinfo {volume} {21}},\
  \bibinfo {pages} {1--19} (\bibinfo {year} {2025})}\BibitemShut {NoStop}%
\bibitem [{\citenamefont {Uzunova}\ \emph {et~al.}(2012)\citenamefont
  {Uzunova}, \citenamefont {Pan}, \citenamefont {Lubchenko},\ and\
  \citenamefont {Vekilov}}]{C2FD20058A}%
  \BibitemOpen
  \bibfield  {author} {\bibinfo {author} {\bibfnamefont {V.}~\bibnamefont
  {Uzunova}}, \bibinfo {author} {\bibfnamefont {W.}~\bibnamefont {Pan}},
  \bibinfo {author} {\bibfnamefont {V.}~\bibnamefont {Lubchenko}}, \ and\
  \bibinfo {author} {\bibfnamefont {P.~G.}\ \bibnamefont {Vekilov}},\
  }\bibfield  {title} {\enquote {\bibinfo {title} {Control of the nucleation of
  sickle cell hemoglobin polymers by free hematin},}\ }\href@noop {} {\bibfield
   {journal} {\bibinfo  {journal} {Faraday Discuss.}\ }\textbf {\bibinfo
  {volume} {159}},\ \bibinfo {pages} {87--104} (\bibinfo {year}
  {2012})}\BibitemShut {NoStop}%
\bibitem [{\citenamefont {Kashchiev}, \citenamefont {Vekilov},\ and\
  \citenamefont {Kolomeisky}(2005)}]{doi:10.1063/1.1943389}%
  \BibitemOpen
  \bibfield  {author} {\bibinfo {author} {\bibfnamefont {D.}~\bibnamefont
  {Kashchiev}}, \bibinfo {author} {\bibfnamefont {P.~G.}\ \bibnamefont
  {Vekilov}}, \ and\ \bibinfo {author} {\bibfnamefont {A.~B.}\ \bibnamefont
  {Kolomeisky}},\ }\bibfield  {title} {\enquote {\bibinfo {title} {Kinetics of
  two-step nucleation of crystals},}\ }\href@noop {} {\bibfield  {journal}
  {\bibinfo  {journal} {J. Chem. Phys.}\ }\textbf {\bibinfo {volume} {122}},\
  \bibinfo {pages} {244706} (\bibinfo {year} {2005})}\BibitemShut {NoStop}%
\bibitem [{\citenamefont {Das}\ \emph {et~al.}(2025)\citenamefont {Das},
  \citenamefont {Zaidi}, \citenamefont {Farag}, \citenamefont {Ruff},
  \citenamefont {Mahendran}, \citenamefont {Singh}, \citenamefont {Gui},
  \citenamefont {Messing}, \citenamefont {Taylor}, \citenamefont {Banerjee},
  \citenamefont {Pappu},\ and\ \citenamefont {Mittag}}]{DAS2025}%
  \BibitemOpen
  \bibfield  {author} {\bibinfo {author} {\bibfnamefont {T.}~\bibnamefont
  {Das}}, \bibinfo {author} {\bibfnamefont {F.~K.}\ \bibnamefont {Zaidi}},
  \bibinfo {author} {\bibfnamefont {M.}~\bibnamefont {Farag}}, \bibinfo
  {author} {\bibfnamefont {K.~M.}\ \bibnamefont {Ruff}}, \bibinfo {author}
  {\bibfnamefont {T.~S.}\ \bibnamefont {Mahendran}}, \bibinfo {author}
  {\bibfnamefont {A.}~\bibnamefont {Singh}}, \bibinfo {author} {\bibfnamefont
  {X.}~\bibnamefont {Gui}}, \bibinfo {author} {\bibfnamefont {J.}~\bibnamefont
  {Messing}}, \bibinfo {author} {\bibfnamefont {J.~P.}\ \bibnamefont {Taylor}},
  \bibinfo {author} {\bibfnamefont {P.~R.}\ \bibnamefont {Banerjee}}, \bibinfo
  {author} {\bibfnamefont {R.~V.}\ \bibnamefont {Pappu}}, \ and\ \bibinfo
  {author} {\bibfnamefont {T.}~\bibnamefont {Mittag}},\ }\bibfield  {title}
  {\enquote {\bibinfo {title} {Tunable metastability of condensates reconciles
  their dual roles in amyloid fibril formation},}\ }\href {\doibase
  https://doi.org/10.1016/j.molcel.2025.05.011} {\bibfield  {journal} {\bibinfo
   {journal} {Molecular Cell}\ } (\bibinfo {year} {2025}),\
  https://doi.org/10.1016/j.molcel.2025.05.011}\BibitemShut {NoStop}%
\bibitem [{\citenamefont {Carnahan}\ and\ \citenamefont
  {Starling}(1969)}]{CarnahanStarling1969}%
  \BibitemOpen
  \bibfield  {author} {\bibinfo {author} {\bibfnamefont {N.~F.}\ \bibnamefont
  {Carnahan}}\ and\ \bibinfo {author} {\bibfnamefont {K.~E.}\ \bibnamefont
  {Starling}},\ }\bibfield  {title} {\enquote {\bibinfo {title} {Equation of
  state for nonattracting rigid spheres},}\ }\href@noop {} {\bibfield
  {journal} {\bibinfo  {journal} {J. Chem. Phys.}\ }\textbf {\bibinfo {volume}
  {51}},\ \bibinfo {pages} {635--636} (\bibinfo {year} {1969})}\BibitemShut
  {NoStop}%
\end{thebibliography}%

\end{document}